\documentclass[12pt]{article}
\usepackage{amsmath,amssymb}

\makeatletter

 \@addtoreset{equation}{section}
\makeatother

\begin{document}
\topmargin -35pt
\oddsidemargin 5mm

\newcommand {\beq}{\begin{eqnarray}}
\newcommand {\eeq}{\end{eqnarray}}
\newcommand {\non}{\nonumber\\}
\newcommand {\eq}[1]{\label {eq.#1}}
\newcommand {\defeq}{\stackrel{\rm def}{=}}
\newcommand {\gto}{\stackrel{g}{\to}}
\newcommand {\hto}{\stackrel{h}{\to}}
\newcommand {\1}[1]{\frac{1}{#1}}
\newcommand {\2}[1]{\frac{i}{#1}}
\newcommand {\thb}{\bar{\theta}}
\newcommand {\ps}{\psi}
\newcommand {\psb}{\bar{\psi}}
\newcommand {\ph}{\varphi}
\newcommand {\phs}[1]{\varphi^{*#1}}
\newcommand {\sig}{\sigma}
\newcommand {\sigb}{\bar{\sigma}}
\newcommand {\Ph}{\Phi}
\newcommand {\Phd}{\Phi^{\dagger}}
\newcommand {\Sig}{\Sigma}
\newcommand {\Phm}{{\mit\Phi}}
\newcommand {\eps}{\varepsilon}
\newcommand {\del}{\partial}
\newcommand {\dagg}{^{\dagger}}
\newcommand {\pri}{^{\prime}}
\newcommand {\prip}{^{\prime\prime}}
\newcommand {\pripp}{^{\prime\prime\prime}}
\newcommand {\prippp}{^{\prime\prime\prime\prime}}
\newcommand {\pripppp}{^{\prime\prime\prime\prime\prime}}
\newcommand {\delb}{\bar{\partial}}
\newcommand {\zb}{\bar{z}}
\newcommand {\mub}{\bar{\mu}}
\newcommand {\nub}{\bar{\nu}}
\newcommand {\lam}{\lambda}
\newcommand {\Lam}{\Lambda}
\newcommand {\lamb}{\bar{\lambda}}
\newcommand {\kap}{\kappa}
\newcommand {\kapb}{\bar{\kappa}}
\newcommand {\xib}{\bar{\xi}}
\newcommand {\ep}{\epsilon}
\newcommand {\epb}{\bar{\epsilon}}
\newcommand {\Ga}{\Gamma}
\newcommand {\rhob}{\bar{\rho}}
\newcommand {\etab}{\bar{\eta}}
\newcommand {\chib}{\bar{\chi}}
\newcommand {\tht}{\tilde{\th}}
\newcommand {\zbasis}[1]{\del/\del z^{#1}}
\newcommand {\zbbasis}[1]{\del/\del \bar{z}^{#1}}
\newcommand {\vecv}{\vec{v}^{\, \prime}}
\newcommand {\vecvd}{\vec{v}^{\, \prime \dagger}}
\newcommand {\vecvs}{\vec{v}^{\, \prime *}}
\newcommand {\alpht}{\tilde{\alpha}}
\newcommand {\xipd}{\xi^{\prime\dagger}}
\newcommand {\pris}{^{\prime *}}
\newcommand {\prid}{^{\prime \dagger}}
\newcommand {\Jto}{\stackrel{J}{\to}}
\newcommand {\vprid}{v^{\prime 2}}
\newcommand {\vpriq}{v^{\prime 4}}
\newcommand {\vt}{\tilde{v}}
\newcommand {\vecvt}{\vec{\tilde{v}}}
\newcommand {\vecpht}{\vec{\tilde{\phi}}}
\newcommand {\pht}{\tilde{\phi}}
\newcommand {\goto}{\stackrel{g_0}{\to}}
\newcommand {\tr}{{\rm tr}\,}
\newcommand {\GC}{G^{\bf C}}
\newcommand {\HC}{H^{\bf C}}
\newcommand{\vs}[1]{\vspace{#1 mm}}
\newcommand{\hs}[1]{\hspace{#1 mm}}
\newcommand{\al}{\alpha}
\newcommand{\be}{\beta}
\newcommand{\bc}{^{\bf C}}
\newcommand{\cu}{{\cal U}}
\newcommand{\csu}{{\cal SU}}
\newcommand{\diag}{{\rm diag.}\,}
\newcommand{\nlsm}{NL$\sigma$Ms~}

\setcounter{page}{0}

\begin{titlepage}

\begin{flushright}
~\\
{\tt hep-th/0312025}\\
December 2003
\end{flushright}
\bigskip

\begin{center}
{\LARGE\bf
Auxiliary Field Methods in Supersymmetric 
Nonlinear Sigma Models
}
\vs{10}

\bigskip
{\renewcommand{\thefootnote}{\fnsymbol{footnote}}
\large\bf Muneto Nitta\footnote{
E-mail: nitta@physics.purdue.edu}
}

\setcounter{footnote}{0}
\bigskip

{\it
Department of Physics, Purdue University, 
West Lafayette, IN 47907-1396, USA
}\footnote{
Address after December 11: 
Department of Physics, Tokyo Institute of 
Technology, Tokyo 152-8551, JAPAN.
}

\end{center}
\bigskip

\begin{abstract}
Auxiliary field methods in $D=2$ (or $3$), ${\cal N}=2$ 
supersymmetric (SUSY) 
nonlinear sigma models (NL$\sig$Ms) are studied.
For these models auxiliary fields as Lagrange multipliers 
belong to a vector or a chiral superfield, 
which gives a K\"ahler quotient of complexified gauge group 
or a holomorphic constraint on it, respectively.
Using these, NL$\sig$Ms on 
all Hermitian symmetric spaces were 
formulated previously. 
In this paper, we formulate new SUSY NL$\sig$Ms 
on some rank-two K\"ahler coset spaces 
as SUSY gauge theories 
with two Fayet-Iliopoulos parameters.

\end{abstract}

\end{titlepage}

\section{Introduction}
The auxiliary field method and the large-$N$ method are 
very powerful tools to study 
non-perturbative effects in a lot of theories, 
such as the Gross-Neveu model,
nonlinear sigma models (NL$\sig$Ms), 
gauge theories, matrix models and so on~\cite{large-N}.
For two-dimensional ($D=2$) \nlsm  on coset spaces $G/H$, 
this method displays 
similarities with four-dimensional ($D=4$) 
gauge theories very easily:  
dynamical mass generation, dynamical symmetry breaking/restoration, 
dynamicaly induced gauge bosons,  
the asymptotic freedom and so on.
Supersymmetric (SUSY) extensions are also 
studied for $D=2$, ${\cal N}=1$ and ${\cal N}=2$ 
SUSY \nlsm\cite{DLD}--\cite{HKNT}. 
$D=3$ \nlsm are non-renormalizable in perturbative method 
but renormalizable in the large-$N$ expansion~\cite{3-dim}. 
SUSY extensions of $D=3$ \nlsm are also studied 
extensively~\cite{3D-SUSY-O(N),3D-SUSY-CPN}.

As an example, 
the auxiliary field method for the $O(N)$ model 
can be illustrated briefly as follows.
Let $g_{ij}(\ph)$ be the metric on $S^{N-1} = O(N)/O(N-1)$ 
with some coordinates $\ph^i$ ($i=1,\cdots,N-1$).
Then the partition function for the $O(N)$ model is given by
\beq
&& Z = \int[d\ph] 
 \exp \left(- \int d^D x 
     \1{2} g_{ij}(\ph) \del_{\mu} \ph^i \del^{\mu} \ph^j\right) \;.
 \label{Z}
\eeq
Classically $O(N)$ symmetry is spontaneously broken down 
to $O(N-1)$.
Introducing an auxiliary field $\sig$ as a Lagrange multiplier, 
this can be rewritten as
\beq
 Z = \int [d\phi d\sig]
 \exp \left[- \int d^D x \left\{
   \1{2} \del_{\mu} \vec{\phi} \cdot \del^{\mu} \vec{\phi} 
  + \sigma(\vec{\phi}^{\,2} - r^2) \right\} \right] \;,
 \label{O(N)}
\eeq
with $\vec{\phi} = \{\phi^A\}$ ($A=1,\cdots,N$) 
being an $O(N)$ vector of scalar fields.
The integration over $\sig$ supplies a constraint 
$\vec{\phi}^{\,2} = r^2$ and 
the Lagrangian in (\ref{Z}) is rederived. 
On the other hand, leaving $\sig$, 
we can first perform the integration 
over dynamical fields $\phi^A$ exactly 
as the Gaussian integral. 
Then, the integration over $\sig$ can be approximated by 
the saddle point in the large-$N$ limit with taking 
$r^2 = N/g^2$. 
We thus obtain the gap equation.  
In $D=2$ $\sig$ gets non-zero vacuum expectation value (VEV) 
$\left<\sig\right> \neq 0$ by solving it. 
Thus the $O(N)$ symmetry is dynamically recovered and 
the mass generation is found as a non-perturbative effect 
contrary to masslessness in the perturbative analysis. 
This agrees with the Coleman's theorem 
which prohibits massless bosons in $D=2$~\cite{Co}.
In $D=3$ there exist two phases of broken and unbroken 
$O(N)$ symmetry.

Therefore the auxiliary field formulation 
is crucial for a study of non-perturbative effects.
It is very easy to find the metric $g_{ij}(\ph)$ 
solving the given constraint on linear fields $\vec{\phi}$:
for instance for the $O(N)$ model, 
putting $\vec{\phi} = (\vec{\ph},\phi^N)$ and 
eliminating the $N$-th component by 
$\phi^N = \sqrt{r^2 - \vec{\ph}^{\,2}}$, 
we obtain the metric 
$g_{ij}(\ph) = \delta_{ij} + 
{\ph^i \ph^j \over r^2 - \vec{\ph}^{\,2}}$ 
expressed in terms of the independent fields $\ph^i$.
It is, however, in general difficult to find 
proper constraints among linear fields $\phi$ 
to give a metric $g_{ij}(\ph)$ of a given model.  
What we really want to do  
for a study of non-perturbative effects 
is in this direction and this is 
the most difficulty of the auxiliary field method. 
In SUSY theories, 
the situation becomes far more complicated
as will be explained below.

For $D=2,3$, ${\cal N}=1$ SUSY NL$\sig$Ms, 
the situation is the same with bosonic case 
because target spaces are Riemannian.
The $D=2$, ${\cal N}=1$ SUSY $O(N)$ model 
was investigated in \cite{WiAl} 
and dynamical chiral symmetry breaking was found. 
The $D=3$, ${\cal N}=1$ SUSY $O(N)$ model 
was discussed in \cite{3D-SUSY-O(N)}.
$D=2,3$, ${\cal N}=2$ SUSY \nlsm  
are obtained as dimensional reduction from 
$D=4$, ${\cal N}=1$ SUSY NL$\sig$Ms. 
For these cases target spaces 
must be K\"ahler~\cite{Zu,WB} 
so that this makes the auxiliary field formulation difficult. 
$D=2$ ($D=3$), ${\cal N}=2$ SUSY NL$\sig$Ms 
on K\"ahler coset spaces $G/H$ may have similarities 
with $D=4$ ($D=5$), ${\cal N}=2$ SUSY QCD~\cite{SUSY-QCD} 
as the same as bosonic theories.
So pursuing similarities of these SUSY models using 
non-perturbative method is very interesting. 

In SUSY theories, 
auxiliary fields as Lagrange multipliers 
belong to superfields.
For these ${\cal N}=2$ SUSY theories, 
there exist vector and chiral superfields 
in terms of $D=4$, ${\cal N}=1$ SUSY~\cite{WB}. 
Using {\it vector} superfields as auxiliary fields, 
the ${\bf C}P^N$ model~\cite{Wi,DDL}  
(${\bf C}P^N = SU(N)/[SU(N-1)\times U(1)]$) 
and the Grassmann model~\cite{Ao} 
($G_{N,M} = U(N)/[U(N-M) \times U(M)]$)
were constructed very long time ago. 

The ${\bf C}P^N$ model is simply given 
in the auxiliary field formulation~\cite{Wi,DDL} as
\beq
 {\cal L} = \int d^4 \theta \, ( e^V \phi\dagg\phi  - c V ) \;,
  \label{CPN}
\eeq
with $\phi$ chiral superfields of 
an $N$-vector, $V$ an auxiliary vector superfield for 
$U(1)$ gauge symmetry, 
and $c$ a real positive constant called 
the Fayet-Iliopoulos (FI) parameter.
In the ${\bf C}P^N$ model, $V$ acquires kinetic term so that
$U(1)$ gauge boson is dynamically induced 
in the large-$N$ limit~\cite{Wi,DDL}. 
In $D=2$ the scalar components in $V$ acquire 
the vacuum expectation value (VEV),  
and there exists the mass generation without breaking 
of the gauge symmetry through the Schwinger mechanism~\cite{Sc}. 
$D=3$, ${\cal N}=2$ SUSY ${\bf C}P^N$ model 
was discussed in Ref.~\cite{3D-SUSY-CPN}.

The $G_{N,M}$ model is given in~\cite{Ao} by
\beq
 {\cal L} = \int d^4 \theta \, 
 \left[ \tr (e^V \Phi\dagg\Phi)  - c\, \tr\, V \right] \;,
  \label{GNM}
\eeq
with $\Phi$ chiral superfields of an $N$ by $M$ matrix 
and $V$ auxiliary vector superfields of an $M$ by $M$ matrix 
for $U(M)$ gauge symmetry. 
This model is expected to induce $U(M)$ gauge bosons 
in the large-$N$ limit.

There was not, however, auxiliary field formulation for 
other $D=2,3$, ${\cal N}=2$ SUSY \nlsm up to a few years ago. 
To overcome such situation, 
the auxiliary field formulation for SUSY \nlsm 
on a broad class of K\"ahler coset spaces, 
Hermitian symmetric spaces (HSS) 
summarized in Table \ref{HSS}, has been given 
in Refs.~\cite{HN1,HN2,HN3}.
Besides auxiliary vector superfields $V$ with 
the K\"ahler potential (\ref{CPN}) or (\ref{GNM}),
we introduced auxiliary {\it chiral} superfields $\sig$ ($\Sig$) 
as summarized in Table \ref{AFF}.
The integration over $V$ gives 
the K\"ahler potential for ${\bf C}P^N$ or $G_{N,M}$, 
whereas the integration over $\sig$ ($\Sig$) 
gives {\it holomorphic} constraints 
which embed the manifold into ${\bf C}P^N$ or $G_{N,M}$.
\begin{table*}[h]
\begin{center}
\begin{tabular}{|c|c|c|c|}
 \noalign{\hrule height0.8pt}
  Type & $G/H$ & complex coordinates $\ph$  
  & $\dim_{\bf C} (G/H)$\\
 \hline
 \noalign{\hrule height0.2pt}
 AIII$_1$&${\bf C}P^{N-1} = {SU(N) \over SU(N-1)\times U(1)}$ 
   & $(N-1)$-vector & $N-1$\\
 AIII$_2$&$G_{N,M} 
           = {U(N) \over U(N-M)\times U(M)}$ 
   & [$M \times (N-M)$]-matrix & $M(N-M)$\\
 BDI     &$Q^{N-2} 
           = {SO(N) \over SO(N-2)\times U(1)}$ 
   & $(N-2)$-vector & $N-2$\\
 CI      & ${Sp(N) \over U(N)}$                          
   & {\small symmetric ($N \times N$)-matrix} & $\1{2}N(N+1)$\\
 DIII    &${SO(2N) \over U(N)}$                         
   & {\small asymmetric ($N \times N$)-matrix} & $\1{2}N(N-1)$\\     
 EIII    &${E_6 \over SO(10)\times U(1)}$                    
   & $16$-spinor & $16$\\
 EVII    &${E_7 \over E_6 \times U(1)}$                      
   & $27$-vector& $27$\\   
 \noalign{\hrule height0.8pt}
 \end{tabular}
 \end{center}
\begin{small}
\caption{Hermitian symmetric spaces (HSS). 
\label{HSS}
}
Classification of Hermitian symmetric spaces (HSS) 
by Cartan, the standard complex coordinates belonging to 
the representation of $H$ and their complex dimensions 
are shown.
\end{small}
\begin{center}
\begin{tabular}{|c||c|c|c||c|c|c|}
 \noalign{\hrule height0.8pt}
  $G/H$ & $\phi$ ($\Phi$) & $\sig$ ($\Sigma$) & $V$
  & superpotential & constraints & hosts \\
 \hline
 \noalign{\hrule height0.2pt}
  ${SO(N) \over SO(N-2)\times U(1)}$ 
      & ${\bf N}$ & ${\bf 1}$  
      & $U(1)$
      & $\sig \phi^2$ 
      & $\phi^2=0$ 
      & ${\bf C}P^{N-1}$\\
  ${SO(2N)\over U(N)}$, ${Sp(N)\over U(N)}$ 
      & $2N \times N$ & $N \times N$ 
      & $U(N)$
      & $\tr (\Sigma \Phi^T J \Phi)$ 
      & $\Phi^T J \Phi = 0$ 
      & $G_{2N,N}$\\
  ${E_6\over SO(10)\times U(1)}$ 
      & ${\bf 27}$ & ${\bf 27}$ 
      & $U(1)$
      & $\Gamma_{ijk}{\sig}^i\phi^j\phi^k$ 
      & $\Gamma_{ijk}\phi^j\phi^k=0$ 
      & ${\bf C}P^{26}$\\
  ${E_7\over E_6 \times U(1)}$ 
      & ${\bf 56}$ & ${\bf 56}$ 
      & $U(1)$
      & $d_{\alpha\beta\gamma\delta}
         {\sig}^{\alpha}\phi^{\beta}
         \phi^{\gamma}\phi^{\delta}$  
      & $d_{\alpha\beta\gamma\delta}\phi^{\beta}
         \phi^{\gamma}\phi^{\delta}=0$ 
      & ${\bf C}P^{55}$\\  
 \noalign{\hrule height0.8pt}
 \end{tabular}
\end{center}
\begin{small}
\caption{Auxiliary field formulation for HSS.\label{AFF}}
Field contents, say dynamical chiral superfields $\phi(\Phi)$ 
and auxiliary chiral and vector superfields $\sig(\Sig)$ and $V$, 
are displayed in the first three rows. 
(We have given the representation of $G$ or matrix sizes 
for $\phi(\Phi)$ and $\sig(\Sig)$ 
and gauge symmetry for $V$.) 
The superpotentials invariant under $G$ and gauge symmetries 
and holomorphic constraints obtained by the integration over 
$\sig(\Sig)$ are also shown. 
In the last row, the host spaces determined by the integration over 
$V$ are shown.  
The second rank symmetric tensor $J$ is given by 
 $J= \left(\begin{array}{cc}{\bf 0}&{\bf 1}_N \cr
       \epsilon{\bf 1}_N &{\bf 0} \end{array}\right)$  
with $\epsilon = +1$ (or $-1$) for $SO(2N)$ (or $Sp(N)$), 
and $\Gamma$ ($d$) is the $E_6$ ($E_7$) symmetric 
third (fourth) rank tensor whose explicit expressions 
can be found in \cite{HN1}. 
\end{small}
\end{table*}

The large-$N$ analysis of these models has become possible. 
The simplest model other than ${\bf C}P^N$ 
is the $Q^N$ model, 
which is called the quadric surface and 
is the coset of $Q^{N-2}=SO(N)/[SO(N-2)\times U(1)]$.
It is given by~\cite{HN1,HKNT}
\beq
 {\cal L} = \int d^4 \theta \, ( e^V \phi\dagg\phi  - c V ) 
 + \left( \int d^2 \theta \, \sigma \phi^2 + {\rm c.c.} \right)
  \label{QN}
\eeq
with $\phi$ dynamical chiral superfields of 
an $N$-vector, $V$ an auxiliary vector superfield for 
$U(1)$ gauge symmetry and 
$\sig$ an auxiliary {\it chiral} superfield.
The integration over $V$ gives the K\"ahler potential 
on ${\bf C}P^N$ and 
the integration over $\sig$ gives 
a holomorphic constraint 
$\phi^2 =0$ among dynamical fields $\phi$. 
So this model is a hybrid of 
the $O(N)$ model (\ref{O(N)}) 
and the ${\bf C}P^N$ model~(\ref{CPN}). 
On the other hand, 
performing the integration over dynamical fields 
$\phi$ in (\ref{QN}) exactly, 
the non-perturbative analysis of the $D=2$ $Q^N$ model 
has been investigated in the large-$N$ method~\cite{HKNT}. 
It has turned out that the $Q^N$ model has 
very interesting features 
which did not exist in the previous models. 
It contains two kinds of non-perturbatively stable vacua; 
one is the Schwinger phase
in which the scalar components of $V$ acquire VEV 
like the ${\bf C}P^N$ model  
and the other is the Higgs phase 
in which the scalar components of $\sig$ get VEV.
The latter is a new vacuum 
for the SUSY \nlsm and 
is interesting if we investigate similarities 
with ${\cal N}=2$ SUSY QCD. 
Both vacua are asymptotically free and 
the mass gap exist due to 
the Shiwinger and Higgs mechanisms. 
$D=3$, ${\cal N}=2$ SUSY $Q^N$ model is also 
discussed recently~\cite{HIT}.

Since the rests of HSS are formulated 
using auxiliary chiral superfields 
besides auxiliary vector superfield, 
these models are also expected to contain 
the Higgs phase besides the Schwinger phase.


Hence our interest is naturally led to 
SUSY \nlsm on more general K\"ahler coset spaces 
other than HSS.
Does dynamical mass generation occur for 
\nlsm on {\it any} K\"ahler coset $G/H$? 
Is any model asymptotically free? 
Which gauge symmetry is dynamically induced 
for a given model?
It is very important to give answers to these questions.
However, before investigating these problems, 
we have to ask if \nlsm on 
{\it any} K\"ahler coset $G/H$ can be formulated 
by the auxiliary field method or not. 
This question is, however, a quite difficult problem.  
In this paper we will make some progress in this problem.

Any K\"ahler coset space can be written as
$G/H = G/[H_{\rm s.s.} \times U(1)^r]$ 
where $H_{\rm s.s.}$ is the semi-simple subgroup in $H$ 
and 
$r \equiv {\rm rank}\, G - {\rm rank}\, H_{\rm s.s.}$ 
is called the rank of this K\"ahler coset space~\cite{Bo}. 
Every HSS is a rank one K\"ahler coset space,
whose Lagrangian has a $U(1)$ or $U(M)$ gauge symmetry 
and one FI parameter 
if formulated by the auxiliary field method.
Other rank one K\"ahler coset spaces 
seem to be relatively easy to be constructed 
in the auxiliary field formulation with 
$U(1)$ or $U(M)$ gauge group~\cite{HN1}. 
In this paper, 
we give the auxiliary field formulation for 
some {\it rank two} K\"ahler coset spaces,
$SU(N)/[SU(N-2) \times U(1)^2]$ and 
$SU(N)/[SU(N-M-L) \times SU(M) \times SU(L) \times U(1)^2]$. 
These models have $U(1)^2$ and $U(M) \times U(L)$ gauge symmetries, 
respectively, and two FI-parameters. 
In addition to auxiliary vector superfields 
for these gauge symmetries, 
some auxiliary chiral superfields are also needed 
even though $G=SU(N)$. 
Non-perturbative studies for these new models become 
possible which will remain as a future work.
We expect that $U(1)^2$ or $U(M) \times U(L)$ 
gauge symmetry is dynamically induced.

This paper is organized as follows.
In Sec.~2, we give
the minimum ingredient of SUSY nonlinear realizations 
needed for later discussions.
Sec.~3 explains how to 
obtain compact K\"ahler coset spaces 
using the super-Higgs mechanism 
eliminating unwanted QNG bosons. 
How this works in the simplest 
${\bf C}P^N$ and $G_{N,M}$ is shown.
In Sec.~4 and 5 we generalize these discussions to 
the rank-two coset spaces 
$SU(N)/[SU(N-2) \times U(1)^2]$ and 
$SU(N)/[SU(N-M-L) \times SU(M) \times SU(L) \times U(1)^2]$, 
respectively. 
Sec.~6 is devoted to summary and discussions.
In Appendix A, we give a review on 
the supersymmetric nonlinear realization 
with K\"ahler $G/H$ focusing on the case of $G=SU(N)$. 
In Appendix B, 
we discuss some geometric aspects 
of these models, 
a relation with some hyper-K\"ahler manifolds 
and an application to construction of a new Calabi-Yau metric.

\section{Supersymmetric Nonlinear Realizations}
The most general discussion for SUSY nonlinear realizations
was discussed by Bando, Kuramoto, Maskawa and Uehara~\cite{BKMU}. 
(For a review see \cite{Ku}.) 
Then they were extensively studied for 
both compact \cite{BFR}--\cite{Ao2} (and references in \cite{BKY}) 
and non-compact \cite{BS}--\cite{FINN} target manifolds. 
In this section we briefly review the minimum of 
SUSY nonlinear realizations needed for 
discussions in the following sections.

A chiral superfield comprises of 
a complex scalar and a Weyl spinor in terms of 
the $D=4$, ${\cal N}=1$ superfield formalism.
In SUSY nonlinear realizations, 
we consider $G\bc$, the complexification of a group $G$, 
because the symmetry group $G$ 
of the superpotential is enhanced to $G\bc$ 
by its holomorphic structure. 
When a global symmetry $G$ is spontaneously 
broken down into its subgroup $H$ 
by vacuum expectation values (VEVs)
with SUSY preserved, 
there appear ordinary Nambu-Goldstone (NG) bosons 
for broken $G$ 
together with additional massless bosons called 
the quasi-NG (QNG) bosons 
for broken $G\bc$ and 
their fermionic SUSY partners~\cite{SNG}.\footnote{
If we assume that all vacuum degeneracy 
come from a symmetry, there are no more massless 
bosons other than these NG and QNG bosons. 
This happens if 
all $G\bc$ invariants composed of fundamental fields 
are fixed to some values. 
So in this case, 
the vacuum manifold becomes a $G\bc$-orbit.
} 
The unbroken subgroup 
of $G\bc$ is also a complex group $\hat H$ 
which contains $H\bc$ as a subgroup. 
Its Lie algebra can be written as\footnote{
We denote the Lie algebra of the group by 
its Calligraphic font.
}
\beq
 \hat {\cal H} = {\cal H}\bc \oplus {\cal B} \;, \label{H-hat}
\eeq
with ${\cal B}$ called the Borel algebra comprising of 
(non-Hermitian) nilpotent generators.
Therefore the vacuum manifold ${\cal M}$ 
parameterized by massless bosons is topologically 
a complex coset space $G\bc/\hat H$, 
which is a non-compact K\"ahler manifold in general.
However we note that the isometry of this coset space 
should be still $G$ but not $G\bc$ because symmetry 
of the K\"ahler potential is not complexified, 
with being different from the superpotential.

A group element in $g \in G\bc$ can be uniquely 
divided into 
$g = \xi \hat h$ with the coset representative 
$\xi$ and $\hat h \in \hat H$. 
Here the coset representative $\xi$ is given by
\beq
 \xi = \exp (\ph \cdot Z) \; \in \; G\bc/\hat H \;
\eeq
with $\ph$ chiral superfields 
and $Z$ complex broken generators in 
${\cal G}\bc - \hat {\cal H}$. 
A set of $Z$ comprises of both 
Hermitian and non-Hermitian generators, 
because $\hat {\cal H}$ in general 
contains non-Hermitian generators in the Borel algebra 
${\cal B}$ as in Eq.~(\ref{H-hat}). 
Corresponding to these Hermitian and 
non-Hermitian broken generators, 
there exist two kinds of massless chiral multiplets, 
called the ``M-type'' and ``P-type'' 
multiplets, respectively. 

In every P-type multiplet, 
both real and imaginary parts of 
a complex scalar field are NG bosons, 
parametrizing compact directions of the manifold ${\cal M}$.
Whereas, in every M-type multiplet, 
one degree of freedom is 
an NG boson and the other is a QNG boson, 
parametrizing a {\it non-compact} direction of ${\cal M}$. 
This can be understood as follows. 
Correspondingly to each 
non-Hermitian broken generator $Z_{\rm P}$, 
there exists a non-Hermitian 
unbroken generator $B$ in 
the Borel algebra ${\cal B}$ 
in $\hat {\cal H}$ (\ref{H-hat})
with $Z_{\rm P}\dagg = B$.
Then, both $X \equiv -i (Z_{\rm P} - B)$ and 
$X' \equiv Z_{\rm P} + B$ are Hermitian.
The coset representative can be transformed to 
an element in its equivalent class as
\beq
 &&\xi = \exp ( \cdots + \ph Z_{\rm P} ) 
  \to \exp ( \cdots + \ph Z_{\rm P} ) \exp ( - \ph^* B ) \non
 && \hs{35}
 = \exp [ \cdots + 
   i ({\rm Re} \ph) X + i ({\rm Im} \ph) X' + O(\ph^2)] \;.
\eeq
Therefore both real and imaginary parts of 
every P-type multiplet $\ph$, 
generated by a non-Hermitian $Z_{\rm P}$, 
parameterize compact directions of ${\cal M}$.
However for every M-type, its real part parameterizes 
a non-compact direction of ${\cal M}$ 
because it is generated by a Hermitian generator.

Let $N_{\rm NG}$, $N_{\rm QNG}$, $N_{\rm P}$ and $N_{\rm M}$ be 
numbers of NG and QNG bosons and P- and M-type multiplets, 
respectively. Then relations
\beq
 N_{\rm NG} = \dim G/H = 2 N_{\rm P} + N_{\rm M} \;, \hs{5} 
 N_{\rm QNG} = N_{\rm M}
\eeq 
hold. 
The complex dimension of the manifold ${\cal M}$ can be 
written as 
\beq
 \dim_{\bf C} {\cal M} = N_{\rm P} + N_{\rm M} = 
 \1{2}(N_{\rm NG} + N_{\rm QNG})\,.
\eeq

If there are no QNG bosons ($N_{\rm QNG} = N_{\rm M} = 0$) 
the manifold becomes compact, whereas if there exists at least one 
QNG bosons ($N_{\rm QNG} = N_{\rm M} \geq 1$) 
the manifold becomes non-compact. 
For non-comact cases, 
NG bosons parameterize a compact homogeneous 
submanifold $G/H$ embedded into 
the total manifold ${\cal M} \simeq G\bc/\hat H$~\cite{KS}. 
For both cases, the isometry of ${\cal M}$ is $G$ but not $G\bc$ 
although $G\bc$ acts on ${\cal M}$ transitively.

Even if the same NG bosons appear for the same $G/H$, 
$N_{\rm QNG}$ depends on the underlying linear model.
Two extreme or natural cases can be considered: 
the case of 
the maximal $N_{\rm QNG}$ equal to $N_{\rm NG}$ 
($N_{\rm P} = 0$) and the case of 
$N_{\rm QNG} = 0$ ($N_{\rm M} =0$).
\begin{enumerate}
\item
If there exist only 
M-type multiplets without P-type multiplets 
($N_{\rm P}=0$, $N_{\rm QNG}=N_{\rm NG}$), 
it is called the ``maximal realization'' 
(or ``fully-doubling''), 
which corresponds to 
$G\bc /\hat H = G\bc/H\bc \simeq T^*(G/H)$.\footnote{
Only the maximal realization cases have a dual description 
by a non-Abelian tensor gauge theory in $D=4$~\cite{FINN}.
} 
Some {\it sufficient} conditions 
for maximal realizations are known~\cite{Le}: 
It occurs 
if a symmetry $G$ is broken down to $H$ 
by VEVs of linear fields 
1) which belong to 
a real representation of $G$
or 
2) with $G/H$ a symmetric space.
In both cases, absence of gauge fields is assumed.

\item
On the other hand, 
when there exist only P-type multiplets without 
M-type multiplets
($N_{\rm M} = N_{\rm QNG} = 0$), 
it is called the ``pure realization'', 
which corresponds to a compact homogeneous K\"ahler manifold 
(K\"ahler coset) $G/H$. 
In this case $G\bc/\hat H \simeq G/H$ holds. 
The K\"ahler metric on arbitrary K\"ahler coset $G/H$ 
was constructed by Borel~\cite{Bo}.  
All K\"ahler coset spaces $G/H$ were classified by 
using the painted Dynkin diagrams~\cite{BFR}.
Their K\"ahler potentials were given by 
Itoh, Kugo and Kunitomo (IKK)~\cite{IKK}.
For pure realizations a no-go theorem due 
to Lerche and Shore is known~\cite{Le,Sh} 
(see also \cite{El,KS,HN1}):
There must appear at least one QNG bosons and 
therefore pure realizations cannot be realized 
if a symmetry $G$ is broken 
by VEVs of linear fields and if
there are no gauge symmetries. 

\end{enumerate}

When we reformulate \nlsm by auxiliary fields, 
dynamical fields must be embedded into fields 
in some linear representations of $G$.
Therefore the Lerche-Shore theorem is the most difficulty 
for the auxiliary field formulation of 
K\"ahler coset spaces $G/H$. 
As discussed in this paper in detail,
this theorem can be avoided 
introducing appropriate gauge symmetry.

\section{Auxiliary Field Formulation for Compact Manifolds}
The ${\bf C}P^N$ model and the $G_{N,M}$ model 
can be easily formulated by 
the auxiliary field method as (\ref{CPN}) and (\ref{GNM}) 
without discussing 
the nonlinear realization method~\cite{Wi,DDL,Ao}. 
The nonlinear realization method 
and the super-Higgs mechanism
play essential roles 
to construct more complicated coset spaces.
We can eliminate QNG bosons by gauging a subgroup of $G$ 
introducing vector multiplets $V$.  
The important thing is that 
vector multiplets $V$ can become massive 
absorbing {\it only M-type} 
chiral multiplets including one NG and one QNG bosons,  
with preserving SUSY.\footnote{
When a gauge field in $V$ absorbs an NG boson in a P-type multiplet, 
SUSY is spontaneously broken~\cite{BW}.
} 
Hence, one $V$ can eliminate one non-compact direction 
with one compact direction. 
If we can eliminate all non-compact directions of QNG bosons, 
a compact manifold of the pure realization can be realized.
Such consideration for 
the ${\bf C}P^N$ model and the $G_{N,M}$ model 
was given in \cite{Ku,HN1} 
and then applied to HSS~\cite{HN1}. 
We briefly review the cases of 
${\bf C}P^N$ and $G_{N,M}$ in this section, 
with some new consideration on 
the geometry of non-compact manifolds before gauging.

\subsection{Auxiliary field formulation for ${\bf C}P^N$}
First, let us discuss ${\bf C}P^{N-1}$.
Prepare linear fields $\phi \in {\bf N}$ of $SU(N)$. 
The system has an additional 
phase symmetry $U(1)_{\rm D}$, given by  
$\phi \to \phi' = e^{i \alpha}\phi$, 
and the total global symmetry becomes 
$G= U(N) = SU(N) \times U(1)_{\rm D}$. 
When the fields $\phi$ develop the VEV 
it can be transformed by $G$ into 
$\left<\phi\right> = (0,\cdots,0,v)^T$ 
with $v$ real positive. 
By this VEV, $G$ is spontaneously 
broken down to its subgroup $H= U(N-1)$.
There appear $N_{\rm NG} = \dim G/H = 2N-1$ NG bosons.

To discuss the whole massless bosons including the QNG bosons,
we consider the complexification of the groups.
The complex unbroken and broken generators are
\beq
 \hat {\cal H} 
= \left( \begin{array}{c|c}
                        & 0      \cr
 {\cal U}(N-1)^{\bf C} & \vdots \cr 
                        & 0      \cr \hline
 {\rm B}\;\;\; \cdots \;\;\; {\rm B} & 0 
 \end{array}\right) \;,\hs{5}
{\cal G}^{\bf C} -\hat {\cal H} 
 = \left( \begin{array}{ccc|c}
        &        &        & {\rm P} \cr
  &{\bf 0}_{N-1} &        & \vdots  \cr 
        &        &        & {\rm P} \cr \hline
      0 & \cdots &  0     & {\rm M} 
 \end{array}\right) \;,\hs{5} \label{gen1}
\eeq
where ${\rm B}$ denote generators in 
a Borel subalgebra ${\cal B}$ in Eq.~(\ref{H-hat}).
Here, M and P denote one Hermitian 
and $N-1$ non-Hermitian 
broken generators, generating M- and P-type 
chiral superfields, respectively 
($N_{\rm M}=1, N_{\rm P} = N-1$).  
So the numbers of NG and QNG bosons are
$N_{\rm NG} = 2N-1$ and $N_{\rm QNG} = 1$, respectively.
The number of the QNG bosons coincides with the number of 
$G$-invariant $|\phi|^2$ as was discussed in \cite{Ni1}.
The manifold $\hat {\cal M}$ can be locally written as 
\beq 
 \hat {\cal M} \simeq {\bf R} \times U(N)/U(N-1)
 \simeq {\bf R} \times SU(N)/SU(N-1)
\eeq 
which is cohomogeneity one.
(From now on we denote non-compact manifolds before gauging 
by $\hat {\cal M}$ and compact manifolds by ${\cal M}$.)

To gauge away unwanted QNG boson 
let us promote $U(1)_{\rm D}$ symmetry to a gauge symmetry 
introducing an auxiliary vector superfield $V$.
The gauge transformation is defined by 
\beq
 \phi \to \phi' = e^{i \Lambda} \phi \;,\hs{5} 
 e^{V} \to e^{V'} 
  = e^{V} e^{-i \Lambda + i \Lambda{}\dagg}\;, 
 \label{gauge-CPN}
\eeq 
with $\Lambda(x,\theta,\thb)$ a gauge parameter of 
a chiral superfield.
Note that this gauge symmetry is complexified 
to $U(1)\bc = GL(1,{\bf C})$ because the scalar component of 
$\Lambda$ is a complex scalar field.
The simplest invariant Lagrangian for these matter 
contents is written as
\beq
 {\cal L} = \int d^4 \theta ( e^V \phi\dagg\phi  - c V ) \;,
\eeq
with $c$ a real positive parameter called 
the Fayet-Iliopoulos (FI) parameter.\footnote{
The most general invariant Lagrangian is 
\beq
 {\cal L} = \int d^4 \theta [ f (e^V \phi\dagg\phi)  - c V ] \;,
\eeq
with an arbitrary function $f$. 
However we can show that we get the same Lagrangian (\ref{CPN-Lag}) 
below when $V$ is eliminated~\cite{HN2}.
}
If we eliminate $V$ by its equation of motion 
$e^V \phi\dagg\phi - c = 0$, 
we obtain 
\beq
 {\cal L} = \int d^4 \theta \, c [\log (\phi\dagg\phi) +1] \;
 = \int d^4 \theta \, c \log (\phi\dagg\phi)\,,
\eeq
where the second term has disappeared 
under the superspace integral.
There still exist the gauge symmetry 
(\ref{gauge-CPN}) for matter fields
in this Lagrangian which can be fixed as 
$\phi^T = (\ph^T,1)$ using $U(1)\bc$.
Then the nonlinear Lagrangian 
in terms of independent fields is obtained as
\beq
 {\cal L} = \int d^4 \theta \, c \log (1+ \ph\dagg\ph) \;.
 \label{CPN-Lag}
\eeq 
We thus have obtained the K\"ahler potential for 
the Fubini-Study metric on ${\bf C}P^{N-1}$.
This construction of ${\bf C}P^N$ is known as 
the K\"ahler quotient method~\cite{LR,HKLR}:
${\bf C}P^{N-1} \simeq \hat {\cal M}/ U(1)\bc$.

Note that $\phi$ can be rewritten as $\phi = \xi v$ with 
$v = (0,\cdots,0,1)^T$ and 
$\xi = \exp 
\left(\begin{array}{cc}{\bf 0}_{N-1} & \ph \cr 
                       {\bf 0} & 0 \end{array}\right) =
\left(\begin{array}{cc}{\bf 1}_{N-1} & \ph \cr 
                      {\bf 0} & 1 \end{array}\right)$ 
being the same as (\ref{CPN-NLR}). 
Comparing the coset generator in $\xi$ and (\ref{gen1}), 
we find that one QNG boson is absorbed, 
with one NG boson,  by the $U(1)$ gauge field $V$ 
and that a pure realization occurs. 

Although we have used the classical equation of motion 
to eliminate $V$ here, we can show that 
this holds in the quantum level 
using the path integral formalism~\cite{HN2,HN3}.

\subsection{Auxiliary field formulation for $G_{N,M}$}
This can be generalized to non-Abelian gauge symmetry
to construct $G_{N,M}$. 
This case is a little bit complicated 
due to non-uniqueness of vacua. 
Let $\Phi$ be an ($N \times M$)-matrix chiral superfield, 
on which the global symmetry 
$G = SU(N) \times SU(M) \times U(1)$ as
\beq
 \Phi \gto \Phi' = g_{\rm L} \Phi g_{\rm R} \;,
\eeq 
with $g_{\rm L} \in SU(N)$ and 
$g_{\rm R} \in U(M) = SU(M) \times U(1)_{\rm D}$.
There exist $M$ independent $G$-invariants composed of 
$\Phi$, given by
\beq
 X_1 \equiv \tr (\Phi\dagg\Phi) \,,\;
 X_2 \equiv \tr [(\Phi\dagg\Phi)^2] \,, \;  \cdots, \; 
 X_M \equiv \tr [(\Phi\dagg\Phi)^M] \, , \label{xs}
\eeq
because the $G$-invariants $\det \Phi\dagg\Phi$ and 
$\tr (\Phi\dagg\Phi)^n$ ($n>M$) are not 
independent with these 
due to the Cayley-Hamilton equation for $M$ by $M$ matrices.
The $G$-invariants (\ref{xs}) determine the cohomogeneity of 
the manifold $\hat {\cal M}$ to be $M$.
The most general invariant Lagrangian is
\beq
 {\cal L} = \int d^4 \theta \; f(X_1,X_2,\cdots,X_M)
\eeq
with an arbitrary function $f$.

Using $G$ symmetry, generic VEVs can be 
transformed into the form of
\beq
 V_{\rm generic} 
   = \left< \Phi \right>
 = \left( \begin{array}{c}
   {\bf 0}_{(N-M)\times M} \cr \hline
     \begin{array}{ccc}
       v_1 &        & {\bf 0} \cr
           & \ddots &  \cr
   {\bf 0} &        & v_M
     \end{array}
 \end{array}\right) \;,\hs{5} 
 \label{GNM-VEV0}
\eeq
with $v_i$ ($i=1, \cdots, M$) $M$ real positive constants. 
These $v_i$'s correspond to (VEVs of) $x_i$'s in Eq.~(\ref{xs})
through $x_j = \sum_{j=1}^M (v_i)^j$.
When all $v_i$'s differ, 
$G$ is spontaneously broken down into 
$H_{\rm g} \equiv SU(N-M) \times U(1)^M$ 
with $i$-th $U(1)$ generated by 
$\diag (\underbrace{1,\cdots,1}_{N-M},
\underbrace{0,\cdots,0}_{i-1},-N+M,
\underbrace{0,\cdots,0}_{M-i})$ in 
$SU(N)$ associated with opposite phase rotation by $U(1)_{\rm D}$.
Hence the the number of NG bosons is 
$N_{\rm NG} = \dim (G/H_{\rm g}) = 2MN - M$. 
The $M$ constants $v_i$ in (\ref{GNM-VEV0}) 
are (VEVs of) QNG bosons 
parametrizing non-compact directions 
${\bf R}^{M}$ in $\hat {\cal M}$, 
and hence $N_{\rm QNG}=M$. 
Then 
$N_{\rm NG} + N_{\rm QNG} = 2MN = 2 \dim_{\bf C} \Phi 
= 2\dim_{\bf C} \hat{\cal M}$ 
correctly holds.
At generic points, 
the manifold can be locally written as  
\beq 
 \hat {\cal M} &\simeq&
 {\bf R}^{M} \times {G / H_{\rm g}} \non
&=& {\bf R}^{M} \times 
 {SU(N) \times U(M) \over SU(N-M) \times U(1)^M }
\simeq 
 {\bf R}^{M} \times 
 {SU(N) \times SU(M) \over SU(N-M) \times U(1)^{M-1} } , \;\;
\eeq
and so it is cohomogeneity $M$. 
In the last equality, the overall phase rotation is cancelled.

When some $v_i$'s coincide, unbroken symmetry is enhanced. 
For instance, when two of them coincide $v_i = v_j$, 
the unbroken symmetry is 
$H' = SU(N-M) \times U(2) \times U(1)^{M-2}$.
So the number of NG bosons $N_{\rm NG}= \dim (G/H') = 2MN - 3$ 
is less than the generic points. 
Some NG bosons have changed into QNG bosons 
with total number of massless bosons unchanged. 
This phenomenon was found by Shore~\cite{KS} and 
was named the ``SUSY vacuum alignment''.
The number of QNG bosons changes from a point to a point, 
but the minimum number of QNG bosons realized at 
the generic points is bounded below by 
the number of the $G$-invariants composed of 
fundamental fields as found in ~\cite{Ni1}. 
So it determines the cohomogeneity of the manifold.

At the most symmetric point 
$v_1 = v_2 = \cdots = v_M \equiv v$, 
the VEV becomes
\beq
 V_{\rm symmetric} 
   = v \left(\begin{array}{cc}{\bf 0}_{(N-M)\times M} \cr
               {\bf 1}_M \end{array} \right)\;. \label{GNM-VEV}
\eeq
The unbroken symmetry is 
$H_0 \equiv SU(N-M) \times SU(M) \times  U(1)$ generated by
\beq
{\cal H} =
 \left(
  \left(\begin{array}{c|c}
 {\cal SU}(N-M)          & {\bf 0}_{(N-M)\times M} \\ \hline
 {\bf 0}_{M\times (N-M)} & A_{M \times M}          
 \end{array}\right) ,
   - A_{M \times M} \right) \oplus \cu(1) \;.
\eeq
with $A_{M \times M}$ $M$ by $M$ generators 
in $SU(M)$. 
Here $U(1)$ is generated by 
$Q_1 \defeq {\rm diag.} \, 
(\underbrace{M,\cdots,M}_{N-M}, 
\underbrace{-N+M,\cdots,-N+M}_{M})$ 
in the $SU(N)$ generators combined with 
the opposite phase rotation by $U(1)_{\rm D}$ 
with fixing the $M \times M$ unit matrix 
in the VEV (\ref{GNM-VEV}).
The number of NG bosons is 
$N_{\rm NG} = \dim (G/H_0) =2MN - M^2$.

At this point complex unbroken and broken generators become 
\beq
 \hat{\cal H} &=& 
 \left(
  \left(\begin{array}{c|c}
 {\cal SU}(N-M)^{\bf C} & {\bf 0}_{(N-M)\times M} \\ \hline
 {\bf B}_{M\times (N-M)}& A_{M \times M}          
 \end{array}\right) ,
   - A_{M \times M} \right) \oplus \cu(1)\bc
\,, \non
 {\cal G}^{\bf C} - \hat{\cal H} &=& 
 \left( \left(\begin{array}{c|c}
 {\bf 0}_{N-M}           & {\bf P}_{(N-M)\times M} \\ \hline
 {\bf 0}_{M\times (N-M)} & {\bf M}_{M \times M}  
 \end{array}\right) , {\bf 0}_{M}  \right) 
  \oplus \cu(1)\bc_{\rm M}\,. \label{GNM-MP}
\eeq
Here ${\bf B}$ represents the matrix of 
the Borel generators 
and ${\bf M}$ and ${\bf P}$ represent 
the matrices of M-type and P-type broken generators, 
respectively, and $U(1)$ in 
${\cal G}^{\bf C} - \hat{\cal H}$ is generated by 
$Q_1$ itself which is also $M$-type. 
So $N_{\rm P} = M(N-M)$ and $N_{\rm M} = M^2$ hold.
The numbers of NG and QNG bosons are 
$N_{\rm NG} = 2N_{\rm P} + N_{\rm QNG} 
= 2MN - M^2$ and $N_{\rm QNG} = M^2$, 
respectively. 
Here $N_{\rm NG}$ is the least and $N_{\rm QNG}$ is the most.
The manifold looks like 
\beq
\hat {\cal M} \simeq
 {\bf R}^{M^2} \ltimes {SU(N) \times U(M) \over SU(N-M) \times U(M)}\;,
\eeq
where we denoted a fiber bundle over a base $B$ with 
a fiber $F$ by $F \ltimes B$. 

Now let us eliminate unwanted QNG bosons with gauging $U(M)$ by 
introducing auxiliary vector superfields $V$
taking a value in ${\cal U}(M)$. 
The $U(M)$ gauge transformation is given by
\beq
 \Phi \to \Phi' = \Phi e^{i \Lambda} \;, \hs{5}
 e^V \to e^{V'} = e^{-i \Lambda} e^V e^{i \Lambda\dagg} \;,
  \label{gauge-tr-GNM}
\eeq
with $\Lambda$ gauge parameters of 
an $M$ by $M$ matrix chiral superfield.
The gauge symmetry is enhanced to its complexification 
$U(M)\bc = GL(M,{\bf C})$. 
Using $SU(N) \times GL(M,{\bf C})_{\rm local}$, 
the VEV can be taken as
\beq
 V_{\rm guage} 
   = \left(\begin{array}{cc}{\bf 0}_{(N-M)\times M} \cr
               {\bf 1}_M \end{array}\right) \;. \label{GNM-gauge-VEV} 
\eeq
Since this VEV takes the form of (\ref{GNM-VEV}) with $v=1$,
breaking pattern is the same with (\ref{GNM-MP}).
The most symmetric point is realized at which 
the number of QNG bosons is the maximum 
$N_{\rm QNG} = M^2$ with coinciding with 
the dimension of the $U(M)$ gauge group.

The simplest invariant Lagrangian is 
\beq
 {\cal L} = \int d^4 \theta \, 
 \left[ \tr (e^V \Phi\dagg\Phi)  - c\, \tr\, V \right] \;,
  \label{GNM-aux}
\eeq
with $c$ real positive.~\footnote{
The most general invariant Lagrangian is
\beq
 {\cal L} = \int d^4 \theta \; 
 f \left(\tr(e^V \Phi\dagg\Phi), \tr[(e^V \Phi\dagg\Phi)^2], 
     \cdots, \tr[(e^V \Phi\dagg\Phi)^M ]
  - c \, \tr\, V \right)
\eeq
with an arbitrary function $f$. 
However we think that 
the resulting Lagrangian after eliminating $V$ 
coincides with the simplest case 
although we do not have a proof.
}
The equation of motion for $V$
\beq
 e^V \Phi\dagg\Phi - c {\bf 1}_M = 0
\eeq
can be used to eliminate $V$. 
Solving this equation as $V = - \log (\Phi\dagg\Phi/c)$ 
and substituting this back into 
the original Lagrangian (\ref{GNM-aux}), we obtain
\beq
 {\cal L} = \int d^4 \theta \, c \log \det (\Phi\dagg\Phi)
\eeq
where a constant has disappeared under the superspace integral.
This still has the $GL(M,{\bf C})$ gauge symmetry (\ref{gauge-tr-GNM}) 
for matter fields.
We can fix this gauge degree of freedom as
\beq 
 \Phi = \left(\begin{array}{cc}
       \ph \cr {\bf 1}_M \end{array}\right)  \label{fixed-Phi}
\eeq  
with $\ph$ an $N-M$ by $M$ matrix of chiral superfields.
Therefore we obtain
\beq
 {\cal L} = \int d^4 \theta \, 
 c \log \det ({\bf 1}_M + \ph\dagg\ph) \;. 
\eeq
This is the K\"ahler potential for the Grassmann manifold
$G_{N,M}$.
In terms of K\"ahler quotient 
we can write $G_{N,M} \simeq \hat {\cal M}/U(M)\bc$.

Note that $\Phi$ in (\ref{fixed-Phi}) can be 
rewritten as $\Phi = \xi V_{\rm gauge}$
using 
$\xi = \exp \left(\begin{array}{cc}  
       {\bf 0}_{N-M} & \ph \cr 
       {\bf 0}_{M \times (N-M)} & {\bf 0}_M \end{array}\right)
=\left(\begin{array}{cc}  {\bf 1}_{N-M} & \ph \cr 
       {\bf 0}_{M \times (N-M)} & {\bf 1}_M \end{array}\right)$ 
in (\ref{GNM-coset})
and $V_{\rm gauge}$ in (\ref{GNM-gauge-VEV}).
Comparing these coset generators and Eq.~(\ref{GNM-MP}), 
we conclude that $M^2$ QNG bosons at 
the most symmetric point are absorbed, 
with the same number of NG bosons,  
by the $U(M)$ gauge fields $V$ 
and that a pure realization occurs. 
The less symmetric does not contribute to 
the resultant manifold.

The $U(M)$ gauge symmetry can be replaced with 
the $U(N-M)$ gauge symmetry with the same 
$SU(N)$ global symmetry considering $\Phi$ an 
$N$ by $N-M$ matrix, 
due to the duality $G_{N,M} \simeq G_{N,N-M}$.

\section{Auxiliary Field Formulation of 
$SU(N)/[SU(N-2)\times U(1)^2]$}

Generalizing the discussion in ${\bf C}P^N$, 
we formulate the rank-two K\"ahler coset space 
$SU(N)/[SU(N-2)\times U(1)^2]$ by the auxiliary field method. 
In the first subsection, we construct a non-compact 
K\"ahler manifold putting a holomorphic constraint 
by an auxiliary chiral superfield without gauging. 
Then in the second subsection, 
we obtain a compact manifold by 
gauging $U(1)^2$ part of the isometry of the non-compact manifold 
introducing auxiliary vector superfields 
with two FI-parameters.

\subsection{Non-compact K\"ahler Manifold}

Let $\phi_1(x,\theta,\thb)$ and $\phi_2(x,\theta,\thb)$ 
being column vectors of chiral superfields, 
belonging to the fundamental 
and the anti-fundamental representations,
${\bf N}$ and $\overline{\bf N}$, of $SU(N)$, respectively. 
They transform under $SU(N)$ as
\beq
 \phi_1 \to \phi_1' = g \phi_1 \;,\hs{5} 
 \phi_2 \to \phi_2' = (g^{-1})^T \phi_2 \;, \label{G-tr}
\eeq
where $g \in SU(N)$ is a matrix element 
of the fundamental representation.\footnote{
$(g^{-1})^T$ is not equivalent to $g^*$ 
when we consider complex extension of the group 
for the symmetry of the superpotential.
We should define the transformation law of 
the anti-fundamental representation by the former.
} 
We put the constraint invariant under $SU(N)$,\footnote{
The constraint $\phi_1 \cdot \phi_2 = a^2$, 
with $a$ being a real constant, was 
discussed in \cite{Ni1}. In this case, one of 
$U(1)$ symmetries (\ref{phi-phase}) 
is explicitly broken. 
} 
\beq
 \phi_1 \cdot \phi_2 = 0 \;. 
\eeq
There exists additional $U(1)^2$ symmetry,  
\beq
 \phi_i \to \phi_i' = e^{i \alpha_i }\phi_i  
  \label{phi-phase}
\eeq
with $\alpha_i$ $(i=1,2)$ being real parameters. 
So the total global symmetry is 
$G \equiv SU(N)\times U(1)^2$.
There exist two $G$-invariants 
$|\phi_1|^2$ and $|\phi_2|^2$ 
so the manifold is cohomogeneity two. 
The most general invariant Lagrangian is given by
\beq
{\cal L} = \int d^4 \theta\, 
 f (\phi_1{}\dagg\phi_1, \phi_2{}\dagg\phi_2) 
 + \left(\int d^2\theta \lam \phi_1 \cdot \phi_2 + {\rm c.c.}\right)\;, 
\eeq
with $f$ an arbitrary function. 
Here $\lam(x,\theta,\thb)$ is an auxiliary chiral superfield
belonging to the singlet of $SU(N)$,
whose $U(1)^2$ transformation is given by
$\lam \to \lam' = e^{- i (\alpha_1 + \alpha_2)} \lam$.

Using the $G$ symmetry, 
the VEVs can be taken as 
\beq
 v_1 = \left<\phi_1\right> 
     = (\underbrace{0,\cdots,0}_{N-1},v)^T \,,\hs{5} 
 v_2 = \left<\phi_2\right> 
     = (\underbrace{0,\cdots,0}_{N-2},u,0)^T \,, 
 \label{VEV}
\eeq
with $v$ and $u$ real positive.
By these VEVs, $G=SU(N) \times U(1)^2$ 
is spontaneously broken down to $H=SU(N-2)\times U(1)^2$. 
Its generators are given by  
\beq
 {\cal H} 
= \left( \begin{array}{c|c|c}
                   & 0 & 0      \cr
   {\cal SU}(N-2)  & \vdots & \vdots \cr 
                              & 0  & 0      \cr \hline
 0 \;\;\;\; \cdots \;\;\;\; 0 & 0  & 0      \cr \hline
 0 \;\;\;\; \cdots \;\;\;\; 0 & 0  & 0 
 \end{array}\right) 
  \oplus \cu(1) \oplus \cu(1)
 \;, 
\eeq
with two $U(1)$'s given by generators
\beq
 Q_1 \equiv \diag (\underbrace{1,\cdots,1}_{N-2},2-N,0), \hs{5}
 Q_2 \equiv \diag (\underbrace{1,\cdots,1}_{N-2},0,2-N) , 
 \label{Q1Q2}
\eeq
in $\csu (N)$
accompanied with 
the $\phi_2$- and $\phi_1$-phase rotations (\ref{phi-phase}) 
with opposite angles, with fixing $v_2$ and $v_1$, respectively.
The number of NG bosons is $N_{\rm NG} = \dim (G/H) = 4N-4$. 

To discuss QNG bosons, 
we consider the action of the complexification of $G$. 
The complex unbroken and broken generators are
\beq
  \hat {\cal H} 
&=& \left( \begin{array}{c|c|c}
                          & {\rm B} & 0      \cr
   {\cal SU}(N-2)^{\bf C} & \vdots  & \vdots \cr 
                          & {\rm B} & 0      \cr \hline
 0 \;\;\;\; \cdots \;\;\;\; 0 &    0    & 0      \cr \hline
{\rm B} \;\;\;\; \cdots \;\;\;\; {\rm B}  & {\rm B} & 0 
 \end{array}\right)  \oplus \cu(1)\bc \oplus \cu(1)\bc  
 \;, \non
  {\cal G}^{\bf C} -\hat {\cal H} 
&=& \left( \begin{array}{ccc|c|c}
        &        &        & 0       & {\rm P} \cr
  &{\bf 0}_{N-2} &        & \vdots  & \vdots  \cr 
        &        &        & 0       & {\rm P} \cr \hline
 {\rm P}& \cdots &{\rm P} & 0 & {\rm P} \cr \hline
      0 & \cdots &  0     & 0       & 0 
 \end{array}\right) 
 \oplus \cu(1)\bc_{\rm M} \oplus \cu(1)\bc_{\rm M}   \;,\hs{5}  
 \label{gen-2M}
\eeq
where ${\rm B}$ denote generators in a Borel subalgebra, 
P denote non-Hermitian
broken generators P-type chiral superfields 
and two $\cu(1)\bc_{\rm M}$'s are given by 
$Q_1$ and $Q_2$ defined in Eq.~(\ref{Q1Q2}), 
both of which generate M-type chiral superfields.
The numbers of the P- and the M-type superfields are
$N_{\rm P} = 2N-3$ and $N_{\rm M} = 2$, respectively. 
Therefore, there appear $N_{\rm NG} = 4N-4$ NG bosons, 
whose number coincides with the dimension of $G/H$, 
and $N_{\rm QNG} =2$ QNG bosons. 
The number of the QNG bosons coincides with 
the number of the $G$ invariants, 
$|\phi_1|^2$ and $|\phi_2|^2$, 
as discussed in \cite{Ni1}. 
The manifold can be locally written as
\beq 
 \hat {\cal M} 
 \simeq {\bf R}^2 \times {SU(N) \times U(1)^2 \over SU(N-2) \times U(1)^2}
 \simeq {\bf R}^2 \times {SU(N) \over SU(N-2)}
\eeq 
which is cohomogeneity two.

\subsection{Gauging: $SU(N)/[SU(N-2)\times U(1)^2]$}
To eliminate unwanted QNG bosons and to 
obtain a compact K\"ahler manifold, 
we promote $U(1)^2$ symmetry (\ref{phi-phase}) to a gauge symmetry
introducing auxiliary vector superfields $V_i(x,\theta,\thb)$ 
$(i=1,2)$. 
The gauge symmetry $U(1)_1 \times U(1)_2$ 
is defined by 
\beq
 \phi_i \to \phi_i' = e^{i \Lambda_i} \phi_i \;,\hs{5} 
 e^{V_i} \to e^{V_i'} 
  = e^{V_i} e^{-i \Lambda_i +i \Lambda_i{}\dagg}\;, 
 \hs{5}
 \lam \to \lam' = \lam e^{-i \Lambda_1 - i \Lambda_2} \,,
 \label{gauge}
\eeq 
with $\Lambda_i(x,\theta,\thb)$ being a chiral superfield as 
a gauge parameter of $U(1)_i$. 
We thus obtain the invariant Lagrangian for 
the auxiliary field formulation of 
$SU(N)/[SU(N-2)\times U(1)^2]$ as
\beq
 {\cal L} = \int d^4 \theta (e^{V_1} \phi_1{}\dagg\phi_1 
 + e^{V_2}\phi_2{}\dagg\phi_2 - c_1 V_1 - c_2 V_2)
 + \left(\int d^2\theta \lam \phi_1 \cdot \phi_2 + {\rm c.c.}\right)\;.
 \label{Lag1}
\eeq

To confirm if this really gives a compact K\"ahler coset space,
we now eliminate auxiliary superfields $\lam$, $V_1$ and $V_2$ 
by their equations of motion.
The integration over $\lam$ gives the holomorphic constraint
\beq
 \phi_1 \cdot \phi_2 = - \chi \cdot \phi + \sig \nu + \rho \mu = 0\;,
 \label{F-con1}
\eeq 
in which we have used the notation
\beq
 \phi_1 = \left(\begin{array}{cc}
            \phi \cr \sig \cr \mu  
          \end{array}\right) \;, \hs{5} 
 \phi_2 = \left(\begin{array}{cc}
            -\chi \cr \nu \cr \rho 
          \end{array}\right) \;, 
\eeq
with $\phi$ and $\chi$ $N-2$ vectors and the rests singlets. 
Introducing a new chiral superfield $\kappa(x,\theta,\thb)$, 
we rewrite this constraint as
\beq
 2 \sig \nu - \chi \cdot \phi = 2 \kappa  \;, \hs{5}
 2 \rho \mu - \chi \cdot \phi = -2 \kappa \;. \label{F-con2}
\eeq
If we eliminate $\kappa$ from these equations, 
we get the constraint (\ref{F-con1}) again. 
Instead, we solve $\rho$ and $\sig$ by other fields:
\beq
 \sig = \1{\nu} \left(\kappa + {\chi \cdot \phi \over 2} \right) 
     , \; \hs{5} 
 \rho = - \1{\mu} \left(\kappa 
    - {\chi \cdot \phi \over 2} \right) \;. \label{F-sol}
\eeq

The integration over $V_i$ gives 
$e^{V_i} \phi_i{}\dagg \phi_i - c_i = 0$, 
which can be solved as 
\beq
 V_i = \log (\phi_i{}\dagg \phi_i/c_i) \;. 
\eeq
Substituting these solutions and (\ref{F-sol}) 
back into the Lagrangian (\ref{Lag1}), 
we obtain 
\beq
 {\cal L} &=& \int d^4\theta  \sum_{i=1}^2 c_i 
 \log (\phi_i{}\dagg \phi_i ) \; \non
 &=& \int d^4 \theta \left[ c_1 \log \left( |\mu|^2 + |\phi|^2 + 
  \left| \1{\nu} \left(\kappa + {\chi \cdot \phi \over 2} \right) 
          \right|^2 \right) 
 \right. \non
 && \left. \hs{10}
 + c_2 \log \left( |\nu|^2 + |\chi|^2 + 
  \left| \1{\mu} \left(\kappa 
    - {\chi \cdot \phi \over 2} \right) \right|^2 \right) 
 \right] \;,
\eeq
where additional constant has disappeared under the integration 
over the superspace.
Since this still has gauge invariance (\ref{gauge}) of 
$U(1)^2$ for matter fields, 
we can take a gauge of $\mu=\nu=1$: 
\beq
 {\cal L} 
 = \int d^4 \theta \left[ 
  c_1 \log \left( 1 + |\phi|^2 + 
  \left|\kappa + {\chi \cdot \phi \over 2}\right|^2 \right) 
 + c_2 \log \left( 1 + |\chi|^2 + 
  \left|\kappa - {\chi \cdot \phi \over 2} \right|^2 \right) 
 \right] \,. 
\eeq
This is the K\"ahler potential of 
$SU(N)/[SU(N-2)\times U(1)^2]$ with 
a suitable complex structure~\cite{BE} 
(see Eq.~(\ref{omegaII}) in Appendix A). 

To see the relation with the non-compact case,
we note the coset representative is given by 
\beq
  \xi = e^{\ph \cdot Z} 
= \left( \begin{array}{c|c|c}
 {\bf 1}_{N-2} & {\bf 0} & \phi   \cr \hline
        \chi^T & 1       & \kappa + \1{2} \chi\cdot \phi \cr \hline
      {\bf 0}  & 0       & 1    
 \end{array}\right) \, \;\;\mbox{ with }\;\;  
\ph \cdot Z
 = \left( \begin{array}{c|c|c}
 {\bf 0}_{N-2} & {\bf 0} & \phi   \cr \hline
        \chi^T & 0       & \kappa \cr \hline
      {\bf 0}  & 0       & 0    
 \end{array}\right) \,   \label{gen2}
\eeq 
where $Z$ represent complex broken generators, 
and $\ph = \{\phi,\chi,\kappa\}$. 
Using this representative, 
the superfields, after solving the constraint and 
fixing a gauge, can be written as  
\beq
 \phi_1 = \xi v_1\,,\hs{5} \phi_2= (\xi^{-1})^T v_2 \,,
\eeq
where $v_i$ are VEVs given 
in (\ref{VEV}) with $u=v=1$.
By comparing the generators (\ref{gen-2M}) and (\ref{gen2}),
it is now obvious that two M-type superfields are 
eliminated by gauging of $U(1)^2$.  From 
the structure of generators (\ref{gen2}), 
we find that this manifold has one of 
two complex structures, called $\Omega_{\rm II}$, 
for $SU(N)/[SU(N-2)\times U(1)^2]$~\cite{BN}.

If we forget the superpotential term 
in the Lagrangian (\ref{Lag1}), 
it becomes the auxiliary field formulation for 
${\bf C}P^{N-1} \times {\bf C}P^{N-1}$. 
Therefore $SU(N)/[SU(N-2)\times U(1)^2]$ 
is embedded into ${\bf C}P^{N-1} \times {\bf C}P^{N-1}$ 
by a holomorphic constraint $\phi_1 \cdot \phi_2=0$ 
so that it is algebraic.

\section{Auxiliary Field Formulation of 
$SU(N)/[SU(N-M-L)\times SU(M) \times SU(L) \times U(1)^2]$}

In this section we show the results of the last section 
can be generalized to 
the auxiliary field formulation for 
$SU(N)/[SU(N-M-L)\times SU(M) \times SU(L) \times U(1)^2]$ 
by promoting gauge groups to non-Abelian groups. 

\subsection{Non-compact K\"ahler Manifold}
Let $\phi_A(x,\theta,\thb)$ ($A=1,\cdots,M$) and 
$\psi_{\alpha}(x,\theta,\thb)$ ($\alpha=1,\cdots,L$) 
be chiral superfields belonging to 
the fundamental and the anti-fundamental representations 
of $SU(N)$, respectively ($N \geq L+M$). 
The transformation law under $SU(N)$ 
is the same as (\ref{G-tr}).
We define matrix chiral superfields 
$\Phi \equiv (\phi_1,\cdots,\phi_M)$ and 
$\Psi \equiv (\psi_1,\cdots,\psi_L)$.  
Additional global symmetries of $U(M)$ and $U(L)$ act from 
the right of $\Phi$ and $\Psi$, respectively, as  
\beq
 && \Phi \to \Phi' = \Phi g_1 \,, \hs{5}  g_1 \in U(M) \,, \non
 && \Psi \to \Psi' = \Psi g_2 \,, \hs{5}  g_2 \in U(L) \,. 
 \label{additional} 
\eeq 
The total global symmetry is $G=SU(N)\times U(M) \times U(L)$. 

We impose $LM$ holomorphic, $G\bc$ invariant 
constraints on these fields: 
\beq
 \psi_{\alpha} \cdot \phi_A = 0 \,, 
 \hs{5} {\rm or}\;\; \Psi^T \Phi = {\bf 0}_{L\times M}  \;.
 \label{const.2}
\eeq
There exist $M+L$ $G$-invariants 
\beq
&& 
 X_1 \equiv \tr (\Phi\dagg\Phi) \,,\;
 X_2 \equiv \tr [(\Phi\dagg\Phi)^2] \,, \;  \cdots, \; 
 X_M \equiv \tr [(\Phi\dagg\Phi)^M] \, , \non
&&
 Y_1 \equiv \tr (\Psi\dagg\Psi) \,,\;
 Y_2 \equiv \tr [(\Psi\dagg\Psi)^2] \,, \;  \cdots, \; 
 Y_L \equiv \tr [(\Psi\dagg\Psi)^L] \, ,
\eeq
so that this manifold is cohomogeneity $M+L$.\footnote{
The invariants made by traces of products 
$M \equiv \Phi\Phi\dagg$ and $N\equiv \Psi\Psi\dagg$ 
are probably not independent of these invariants 
although we do not have a proof. 
It is plausible considering the relation with VEVs as below.
}
The most general invariant Lagrangian is
\beq
 {\cal L} = \int d^4\theta 
  f (X_1, \dots X_M, Y_1,\cdots,Y_L ) 
 + \left[ \int d^2\theta \, 
        \tr (\lam \Psi^T \Phi) + {\rm c.c.} \right] \,,
\eeq
with $\lam$ an $M \times L$ 
matrix of auxiliary chiral superfields 
$\lam_{A \alpha} (x,\theta,\thb)$ 
of Lagrange multipliers
for the constraints (\ref{const.2}). 

Using $G=SU(N)\times U(M) \times U(L)$, 
the generic VEVs can be transformed into 
\beq
&& 
 V_1 = \left< \Phi \right> 
 = \left(\begin{array}{c}
   {\bf 0}_{(N-M)\times M} \cr \hline
   \begin{array}{ccc}
       v_1 &        & {\bf 0} \cr
           & \ddots &  \cr
   {\bf 0} &        & v_M
     \end{array}  \cr
 \end{array}\right) \,, \non
&& 
 V_2^T 
 = \left< \Psi \right>^T
 = \left( \begin{array}{c|c|c}
   {\bf 0}_{L \times (N-M-L)}    &
       \begin{array}{ccc}
           u_1 &        & {\bf 0} \cr
               & \ddots &  \cr
       {\bf 0} &        & u_L
       \end{array}                 & 
  {\bf 0}_{L \times M} 
 \end{array} \right) \,.
\eeq
with $v_i$ ($i=1,\cdots,M$) and 
$u_a$ ($a=1,\cdots,L$) $M+L$ 
real positive constants. 
These VEVs are related with VEVs of the invariants 
$X_i$ and $Y_a$ through
$X_i = \sum_{j=1}^M (v_j)^i$ and 
$Y_a = \sum_{b=1}^L (u_b)^a$. 
$G$ is spontaneously broken down to
$H_{\rm g} = SU(N-M-L) \times U(1)^{M+L}$ 
and $N_{\rm NG} = \dim (G/H_{\rm g}) = 2(NM + LN - LM) - M - L$. 
The number of the QNG bosons is $N_{\rm QNG} = M+L$. 
So equations $N_{\rm NG} + N_{\rm QNG} 
= 2 (NM + LN - LM) 
= 2 (\dim_{\bf C} \Phi + \dim_{\bf C} \Psi - LM) 
= 2 \dim_{\bf C} \hat {\cal M}$ 
correctly holds. 
At generic points, 
the manifold can be locally written as  
\beq 
 \hat {\cal M} &\simeq&
 {\bf R}^{M+L} \times {G / H_{\rm g}}  
 = {\bf R}^{M+L} \times 
 {SU(N) \times U(M) \times U(L) \over 
  SU(N-M-L) \times U(1)^{M+L} }  \non
 &\simeq& {\bf R}^{M+L} \times 
 {SU(N) \times SU(M) \times SU(L) \over 
  SU(N-M-L) \times U(1)^{M+L-2} } \;,
\eeq
so that it is cohomogeneity $M+L$.

At the most symmetric points, 
the VEVs become 
\beq
 V_1 = \left< \Phi \right> 
 = v \left(\begin{array}{c}
   {\bf 0}_{(N-M)\times M} \cr
   {\bf 1}_M \cr
 \end{array}\right) \,, \hs{5} 
 V_2^T 
 = \left< \Psi \right>^T
 = u \left( {\bf 0}_{L \times (N-M-L)},
              {\bf 1}_L, {\bf 0}_{L \times M} \right) \,.
\eeq
The unbroken subgroup is 
$H_0 = SU(N-M-L) \times SU(M) \times SU(L) \times U(1)^2$ 
whose generators are given by 
\beq
&&{\cal H} = \left(\
  \left(\begin{array}{c|c|c}
 {\cal SU}(N-M-L)  & {\bf 0}_{(N-M-L)\times L}  
                   & {\bf 0}_{(N-M-L)\times M} \\ \hline
 {\bf 0}_{L \times (N-M-L)} & X_{L\times L} & {\bf 0}_{L\times M}\\ \hline
 {\bf 0}_{M\times (N-M-L)}  & {\bf 0}_{M \times L}& Y_{M\times M}      
 \end{array}\right) ,
 -X_{L\times L}, -Y_{M\times M} \right) \non
 && \hs{15} \oplus \cu(1) \oplus \cu(1)  ,
\eeq
with $X$ and $Y$ generators of 
${\cal SU}(L)$ and ${\cal SU}(M)$, respectively.
Here the two $U(1)$'s are given by 
\beq 
&& Q_1 \equiv \diag (\underbrace{L,\cdots,L}_{N-M-L},
\underbrace{-N+M+L,\cdots, -N+M+L}_L,\underbrace{0,\cdots,0}_M), \non
&& Q_2 \equiv \diag (\underbrace{M,\cdots,M}_{N-M-L}, 
\underbrace{0,\cdots,0}_L,
\underbrace{-N+M+L,\cdots, -N+M+L}_M)  \label{Q1Q2-GNM}
\eeq 
in ${\cal SU}(N)$ combined with
phase rotations with opposite angles 
by $U(1)_1$ and $U(1)_2$, respectively.

Complex unbroken and broken generators become like 
\beq
&& \hat{\cal H} = 
 \left(
  \left(\begin{array}{c|c|c}
   {\cal SU}(N-M-L)^{\bf C} & {\bf B}_{(N-M-L)\times L}  
                            & {\bf 0}_{(N-M-L)\times M} \\ \hline
 {\bf 0}_{L \times (N-M-L)} & X_{L\times L} & {\bf 0}_{L\times M}\\ \hline
  {\bf B}_{M\times (N-M-L)} & {\bf B}_{M \times L}& Y_{M\times M},    
 \end{array}\right), 
  -X_{L\times L}, -Y_{M\times M}  \right) \non
 && \hs{15} \oplus \cu(1)\bc \oplus \cu(1)\bc
 \,, \non
 &&
 {\cal G}^{\bf C} - \hat{\cal H} = 
 \left( \left(\begin{array}{c|c|c}
  {\bf 0}_{N-M-L} & {\bf 0}_{(N-M-L)\times L} 
     & {\bf P}_{(N-M-L)\times M} \\ \hline
  {\bf P}_{L \times (N-M-L)} & {\bf M}_{L\times L} & {\bf P}_{L\times M}\\ \hline
 {\bf 0}_{M\times (N-M-L)}& {\bf 0}_{M \times L}& {\bf M}_{M\times M} 
 \end{array}\right) ,
 {\bf 0}_{L \times L}, {\bf 0}_{M \times M}
 \right)  \non
 && \hs{25} \oplus \cu(1)\bc_{\rm M} \oplus \cu(1)\bc_{\rm M} \,,
\eeq
respectively, 
where each subscript denotes the size of each block, 
and two $U(1)$'s in ${\cal G}^{\bf C} - \hat{\cal H}$ 
are generated by $Q_1$ or $Q_2$ in Eq.~(\ref{Q1Q2-GNM}), 
both of which are M-types.  
The numbers of the M- and P-type superfields are 
$N_{\rm M} = L^2 + M^2$ and 
$N_{\rm P} = M(N-M) + L(N-M-L)$, respectively. 
There appear $N_{\rm NG} = 2(NM + NL - L) - M^2 -L^2$ NG bosons 
and $N_{\rm QNG} = (L^2 + M^2)$ QNG bosons. 
The manifold looks like
\beq
\hat {\cal M} \simeq
 {\bf R}^{M^2+L^2} 
 \ltimes {SU(N) \times U(M) \times U(L) 
         \over SU(N-M-L) \times U(M) \times U(L)}\,.
\eeq

\subsection{Gauging: $SU(N)/[SU(N-M-L)\times SU(M) 
\times SU(L) \times U(1)^2]$}
To absorb M-type superfields,  
we promote the global symmetries of (\ref{additional}) to 
gauge symmetries, by introducing auxiliary vector superfields  
$V_1$ and $V_2$ of $U(M)$ and $U(L)$ gauge fields, respectively.
These $U(M) \times U(L)$ gauge symmetries are defined by 
\beq
 && \Phi \to \Phi' = \Phi e^{i \Lam_1} \,, \hs{5} 
    \Psi \to \Psi' = \Psi e^{i \Lam_2} \;, \non
 && e^{V_1} \to e^{V_1'} 
    = e^{-i \Lam_1} e^{V_1} e^{i \Lam_1\dagg}\,,\hs{5}
   e^{V_2} \to e^{V_2'} 
    = e^{-i \Lam_2} e^{V_2} e^{i \Lam_2\dagg} \,, \non
 && \lam \to \lam' = e^{-i \Lam_1} \lam e^{-i \Lam_2^T} \,,
 \label{gauge2}
\eeq
where $\Lam_1(x,\theta,\thb)$ and $\Lam_2(x,\theta,\thb)$ 
are chiral superfields of gauge parameters, 
taking values in ${\cal U}(M)$ and ${\cal U}(L)$, respectively.  
Then the Lagrangian can be written as 
\beq
 &&{\cal L} = \int d^4\theta \left[ \tr(\Phi\dagg\Phi e^{V_1})
 + \tr(\Psi\dagg\Psi e^{V_2}) - c_1 \tr V_1 - c_2 \tr V_2 \right] \non
 && \hs{10}
 + \left[ \int d^2\theta \tr (\lam \Psi^T \Phi) + {\rm c.c.} \right] \,. 
 \label{Lag2}
\eeq

We decompose the matrix chiral superfields into 
submatrices, like
\beq
 \Phi = \left(\begin{array}{c}
  \ph_{IA} \\
  \sig_{\alpha A} \\
  \mu_{AB}
 \end{array}\right) \,, \hs{5} 
 \Psi = \left(\begin{array}{c}
  -\chi_{I\alpha} \\
  \nu_{\alpha \beta} \\
  \rho_{A\alpha}
 \end{array}\right) \,,
\eeq
where a new index $I$, which runs from $1$ to $N-M-L$, 
has been introduced.
In terms of these decompositions, 
the integration over $\lam_{\alpha A}$ yields 
the $LM$ holomorphic constraints 
\beq
 \psi_{\alpha} \cdot \phi_A
 = - \chi^T_{\al I} \ph_{IA} + \nu^T_{\al\be}\sig_{\be A}
  + \rho^T_{\al B} \mu_{BA}  = 0 \,, \label{const.mat.}
\eeq
in which the summation over repeated indices is implied. 
Introducing a new $L\times M$ matrix chiral superfield, 
$\kappa_{\al A}(x,\theta,\thb)$, 
these constraints can be rewritten as
\beq
 2 \nu^T_{\al\be}\sig_{\be A} - \chi^T_{\al I} \ph_{IA}
  &=& 2 \kappa_{\al A} \,,\non
 2 \rho^T_{\al B} \mu_{BA} - \chi^T_{\al I} \ph_{IA} 
  &=& - 2 \kappa_{\al A} \,.
\eeq
Elimination of $\kappa$ from these equations gives 
the original constraints (\ref{const.mat.}). 
Instead, we express $\sig$ and $\rho$ by other fields 
in the region such that 
$\det \mu \neq 0$ and $\det \nu \neq 0$ hold: 
\beq
 \sig = \nu^{-1 T} 
       \left( \kappa + \1{2} \chi^T \ph \right)\,,\hs{5}
 \rho = - \mu^{-1 T} \left( \kappa - \1{2} \ph^T \chi \right)\,, 
 \label{F-sol-mat.}
\eeq
where we have used the matrix notation. 

The equations of motion for $V_1$ and $V_2$ read 
$e^{V_1}\Phi\dagg\Phi - c_1 {\bf 1}_M = 0$ and 
$e^{V_2}\Psi\dagg\Psi - c_2 {\bf 1}_L = 0$, respectively.  
These equations can be solved, to give 
\beq
 V_1 = - \log \det \left({\Phi\dagg\Phi \over c_1}\right) \,,\hs{5} 
 V_2 = - \log \det \left({\Psi\dagg\Psi \over c_2}\right) \,.
\eeq
Substituting these solutions back into 
the Lagrangian (\ref{Lag2}), we obtain 
\beq
 {\cal L} 
 &=& \int d^4 \theta \left[ 
   c_1 \log \det \Phi\dagg\Phi 
 + c_2 \log \det \Psi\dagg\Psi\right] \,\non
 &=& \int d^4 \theta \left[ 
   c_1 \log \det (\ph\dagg\ph + \sig\dagg\sig + \mu\dagg\mu) 
 + c_2 \log \det (\chi\dagg\chi + \nu\dagg\nu + \rho\dagg\rho) 
\right] \, .
\eeq
By substituting (\ref{F-sol-mat.}) into this Lagrangian  
and taking a gauge fixing of  
$\mu = {\bf 1}_M$ and $\nu = {\bf 1}_L$, we obtain 
\beq
 &&{\cal L} 
 = \int d^4 \theta \left[ 
   c_1 \log \det \left\{ {\bf 1}_M + \ph\dagg\ph 
          + \left( \kappa\dagg + \1{2} \ph\dagg \chi^* \right) 
            \left( \kappa + \1{2} \chi^T \ph \right) \right\} 
 \right.\non 
 &&\left. \hs{18}
 + c_2 \log \det \left\{ {\bf 1}_L + \chi\dagg\chi 
          + \left( \kappa\dagg - \1{2} \chi\dagg \ph^*  \right)
            \left( \kappa - \1{2} \ph^T \chi \right)
 \right\} 
\right] \, . \label{result2}
\eeq
We thus have obtained the K\"ahler potential 
of $SU(N)/[SU(N-M-L)\times U(M) \times U(L)]$ 
with a suitable complex structure 
through a K\"ahler quotient: 
${\cal M} \simeq \hat {\cal M}/ [U(M)\bc \times U(L)\bc]$

Since the Lagrangian (\ref{Lag2}) becomes 
the auxiliary field formulation for 
$G_{N,M} \times G_{N,L}$
throwing away its superpotential, 
we find that $SU(N)/[SU(N-M-L)\times U(M) \times U(L)]$ is 
embedded into $G_{N,M} \times G_{N,L}$ 
by holomorphic constraints by $\Psi^T \Phi = {\bf 0}_{L\times M}$.

An interesting thing is that 
there exists the triality between theories with
the same $U(N)$ flavor symmetry and 
following three different gauge groups: 
$U(M) \times U(L)$, 
$U(N-M-L) \times U(L)$ and 
$U(M) \times U(N-M-L)$.

\section{Summary and Discussions}

We have given the auxiliary field formulation 
for $D=2,3$, ${\cal N}=2$ 
SUSY NL$\sig$Ms on the rank two K\"ahler coset spaces 
$SU(N)/[SU(N-2)\times U(1)^2]$ and 
$SU(N)/[SU(N-M-L) \times SU(M) \times SU(L) \times U(1)^2]$ 
as $U(1)^2$ and $U(M)\times U(L)$ gauge theories with 
the Lagrangians (\ref{Lag1}) and (\ref{Lag2}), 
respectively. 
In addition to auxiliary vector superfields $V_1$ and $V_2$ 
for these gauge groups, 
we have also needed auxiliary chiral superfields $\lam$ 
to give holomorphic constraints among two irreducible 
representations. 
For both cases the Lagrangian includes 
two FI parameters $c_1$ and $c_2$ 
which represent 
two free parameters (decay constants) of 
the resultant K\"ahler coset spaces.
Non-perturbative analyses using the large-$N$ method 
for these new models in $D=2,3$ dimensions 
have become possible, 
which remains as a future work.

Let us discuss possibility to construct more general models 
by the auxiliary field method.
First of all 
the K\"ahler coset spaces discussed in this paper 
have particular complex structure 
$\Omega_{\rm II}$ (see Appendix A). 
However, in general, the same rank-two coset space 
allows two inequivalent complex structures 
$\Omega_{\rm I}$ and $\Omega_{\rm II}$ 
as discussed in Appendix A. 
So the question is whether it is possible to 
construct a model with another complex 
structure $\Omega_{\rm I}$ or not.
In the case of $SU(N)/[SU(N-2)\times U(1)^2]$ with 
$\Omega_{\rm I}$, it is natural to prepare 
the two sets of fields 
$\phi_1$ and $\phi_2$ both belonging to 
the fundamental representation ${\bf N}$ 
of $SU(N)$, instead of ${\bf N}$ and $\overline{\bf N}$ 
for $\Omega_{\rm II}$. 
Actually, 
the authors in Ref.~\cite{IOR}  
constructed a {\it bosonic} NL$\sig$M on 
$SU(N)/[SU(N-2)\times U(1)^2]$ with 
the complex structure $\Omega_{\rm I}$
by the auxiliary field method. 
They introduced an $U(2)$ gauge symmetry 
explicitly broken by two constraints 
$\phi_1{}\dagg\phi_1 = c_1$ and $\phi_2{}\dagg\phi_2 = c_2$ 
with $c_1 \neq c_2$. 
These constraints can be embedded into bosonic parts of 
the SUSY $U(1)^2$ gauge theory 
with two FI-parameters $c_1$ and $c_2$ in the Wess-Zumino gauge.
They moreover imposed an additional constraint 
$\phi_1{}\dagg\phi_2 = 0$ 
instead of $\phi_1 \cdot \phi_2 = 0$ for $\Omega_{\rm I}$. 
It is, however, difficult to embed this constraint 
to a SUSY theory.

Second, let us discuss rank-two coset spaces $G/H$ with 
other groups $G$.
For rank-one case it was possible to reduce 
$G=SU(N)$ to other groups $G' \subset G$ imposing 
$G'$-invariant F-term (holomorphic) constraints 
by auxiliary chiral superfields 
which give an embedding for the whole coset $G'/H' \subset G/H$, 
as in the Lagrangian (\ref{QN}) for 
$Q^{N-2} = SO(N)/[SO(N-2)\times U(1)] \subset {\bf C}P^{N-1}$. 
In principle, it should work also for rank-two cases 
but naive attempts failed. 
This remains for a future work.

Third, we would like to discuss higher-rank coset spaces. 
The Lerche-Shore theorem \cite{Le,Sh} implies 
that any K\"ahler $G/H$ needs a gauge group 
to be formulated by linear fields. 
It is natural for a gauge group to 
include the $U(1)^r$ factor with $r$ FI-parameters 
for a rank-$r$ coset space.
Moreover we should introduce at least 
$r$ irreducible representations of $G$
and put suitable F-term constraints among them 
which are needed to fix all 
$G\bc$-invariants composed of them.
However the similar problem for $SU(N)/[SU(N-2)\times U(1)^2]$ 
with $\Omega_{\rm I}$ occurs and 
we are unable to achieve this in the present time.

Perturbatively, dynamics of 
$D=2$, ${\cal N}=2$ SUSY NL$\sig$Ms 
have very different features according to 
their first Chern classes $c_1(M)$
on the target manifold $M$. 
Calabi-Yau manifolds $M$ have vanishing  first Chern classes 
$c_1(M) = 0$. 
$D=2,{\cal N}=2$ SUSY NL$\sig$Ms on these manifolds 
are conjectured to be finite to all orders 
and are considered as models 
of superstring theory~\cite{finite}. 
On the other hand, 
all K\"ahler coset spaces have positive 
first Chern classes: $c_1(G/H) > 0$. 
For all NL$\sig$Ms on $M$ with $c_1 (M) >0$, 
it is conjectured that these models are asymptotically free and 
have the mass gap like $D=4$ QCD.

Non-perturbative analyses for 
these features were discussed extensively 
using the mirror symmetry~\cite{HV}.
Exact beta functions for $D=2$ NL$\sig$Ms on Hermitian 
symmetric spaces were derived using 
the instanton method~\cite{exact-beta}.
$D=2,3$, ${\cal N}=2$ SUSY \nlsm were also 
discussed using the Wilsonian renormalization group (WRG)~\cite{HI}, 
in which they used the so-called K\"ahler 
normal coordinates~\cite{KNC} to derive the WRG equation.
In particular, $D=2,3$, ${\cal N}=2$ SUSY \nlsm on 
the K\"ahler-Einstein manifolds including 
K\"ahler coset spaces were discussed.

Combined with these several methods, 
we expect that the large-$N$ method plays an important role 
to reveal non-perturbative aspects of SUSY \nlsm 
and their similarity with $D=4,5$ QCD.

\section*{Acknowledgements} 

The author thanks Kiyoshi Higashijima and Makoto Tsuzuki 
for a collaboration in early stages of this work. 
His work is supported by the U.~S. Department 
of Energy under grant DE-FG02-91ER40681 (Task B).

\begin{appendix}
\section{Pure realizations for $G=SU(N)$ cases}
We work out theories with the global symmetry $SU(N)$. 
For details see the original Refs.~\cite{BKMU,IKK} 
or the reviews \cite{BKY,Ku}.
Pure realizations with $G=SU(N)$ occur 
if and only if $H$ is the form of 
$H = \prod_{i=1}^{r+1} SU(n_i) \times U(1)^r$ 
with $\sum_{i=1}^{r+1} n_i = N$ 
and $r (\geq 1)$ called the rank of 
this K\"ahler coset space $G/H$: 
\beq
 {\cal H} =
 \left( \begin{array}{c|c|c|c}
  \csu(n_1) &         &        &              \cr \hline
            &\csu(n_2)&        &              \cr \hline
            &         & \ddots &              \cr \hline
            &         &        & \csu(n_{r+1})  
 \end{array}\right)  
 \oplus \underbrace{\cu(1) \oplus \cdots \oplus \cu(1)}_r \;.
\eeq
All off-diagonal blocks are zero matrices and 
$r$ $U(1)$ generators $Q_{\alpha}$ $(\alpha=1,\cdots,r)$ 
are given by 
\beq 
Q_{\alpha} = \diag 
(\underbrace{n_{\alpha+1}, \cdots , n_{\alpha+1}}_{n_1}, 
 \underbrace{0,\cdots,0}_{\sum_{\beta=2}^{\alpha} n_{\beta}}, 
 \underbrace{- n_1,\cdots,-n_1}_{n_{\alpha+1}}
 \underbrace{0,\cdots,0}_{\sum_{\beta=\alpha+2}^{r+1} n_{\beta}}).
\eeq

A complex structure on $G/H$ is defined as follows.
Taking linear combination of $U(1)$ generators $Q_{\alpha}$ 
in ${\cal H}$, we define the $Y$-charge by 
\beq 
 Y = \sum_{\alpha=1}^r c_{\alpha} Q_{\alpha} \label{Y}
\eeq 
with $c_{\alpha} \in {\bf R}$. 
Complex linear combination $\sum_I C_I X_I$ ($C_I \in {\bf C}$)
of the real coset generators 
$X_I \in {\cal G} - {\cal H}$ can be divided into 
$B_i \in \hat{\cal H}$ and 
$Z_i \in {\cal G}\bc - \hat{\cal H}$  
according to positive and negative $Y$-charges, respectively: 
$[Y,B_i] \sim + B_i$ and $[Y,Z_i] \sim - Z_i$. 
Note that the generators in ${\cal H}$ carry zero $Y$-charges 
and therefore 
all generators in $\hat {\cal H}$ carry non-negative $Y$-charges. 
Thus, rank one coset spaces have one complex structure.
Rank (more than) two coset spaces have (more than) two
inequivalent complex structures.
The complex coset representative can be defined by
$\xi = \exp (\ph \cdot Z) \in G\bc/\hat H$ with 
$\ph^i$ NG chiral multiplets 
whose scalar components are both genuine NG bosons. 
There exists homomorphism $G\bc/\hat H \simeq G/H$ 
for pure realizations because there are no QNG bosons.

Once a complex structure is provided we can construct 
a $G$-invariant K\"ahler potential on $G/H$.
There exist $r$ 
projection operators $\eta_{\alpha}$ ($\alpha=1,\cdots,r$) 
satisfying the conditions
\beq
 \eta \hat H \eta = \hat H \eta, \hs{5} 
 \eta^2 = \eta,  \hs{5}
 \eta\dagg = \eta \, 
\eeq 
in the space of the fundamental representation ${\bf N}$ of $G$.
We can take these $\eta$ according to the $Y$-charges of 
the fundamental representation space as follows: 
first decompose the fundamental representation of $G$ 
into $H$-irreducible sectors.
Second take the $\alpha$-th projection 
$\eta_{\alpha}$ to project out 
the first $\alpha$ sectors with lower $Y$-charges 
from the lowest $Y$-charge sector.
Using these projections, the K\"ahler potential on $G/H$ is given by
\beq
 K = \sum_{\alpha=1}^r c_{\alpha} \log \det{}_{\eta} \xi\dagg \xi \;,
\eeq
with $\xi \in G\bc/\hat H$ in the fundamental representation 
and $\det_{\eta}$ denoting the determinant in 
the subspace projected by $\eta$.
Here $c_{\alpha}$ can be shown to coincide with 
the coefficients in the $Y$-charge (\ref{Y})~\cite{IKK}.

We give the following four examples discussed 
in this paper: 
1) ${\bf C}P^{N-1} = SU(N)/[SU(N-1)\times U(1)]$,
2) $G_{N,M} = SU(N)/[SU(N-M) \times SU(M) \times U(1)]$,
3) $SU(N)/[SU(N-2)\times U(1)^2]$ and
4) $SU(N)/[SU(N-M-L) \times SU(M) \times SU(L) \times U(1)^2]$.

\medskip

1) ${\bf C}P^{N-1} = SU(N)/[SU(N-1)\times U(1)]$\\
This is the simplest example called the projective space.
If we define the $Y$-charge by $Y \equiv - Q_1$
with 
$Q_1 \equiv {\rm diag.}\, 
(\underbrace{1,\cdots,1}_{N-1},-N+1) \in {\cal H}$,
complex unbroken and broken generators, 
carrying positive and negative (or zero) $Y$-charges, 
are found to be
\beq
 \hat {\cal H} 
= \left( \begin{array}{c|c}
                   & 0      \cr
 {\cal SU}(N-1)\bc & \vdots \cr 
                   & 0      \cr \hline
 {\rm B}\;\;\; \cdots \;\;\; {\rm B} & 0
 \end{array}\right) 
 \oplus \cu(1)\bc \;,\hs{5}
{\cal G}^{\bf C} -\hat {\cal H} =
\left( \begin{array}{ccc|c}
        &        &        & {\rm P} \cr
  &{\bf 0}_{N-1} &        & \vdots  \cr 
        &        &        & {\rm P} \cr \hline
      0 & \cdots &  0     & 0         
 \end{array}\right),  
\eeq
respectively, with $\cu(1)\bc$ generated by $Q_1$.
Then the complex coset representative is given by
\beq
 \ph \cdot Z = 
 \left(\begin{array}{cc}
  {\bf 0}_{N-1} & \ph \cr
       {\bf 0}  & 0 \end{array} \right)\;,\hs{5} 
 \xi = \left(\begin{array}{cc}
  {\bf 1}_{N-1} & \ph \cr
       {\bf 0}  & 1 \end{array} \right)\label{CPN-NLR}
\eeq
with $\ph$ an ($N-1$)-vector of $SU(N-1)$.
The projection operator is 
$\eta = \diag (\underbrace{0,\cdots,0}_{N-1},1)$ and 
the K\"ahler potential is 
\beq
 K = c \log \det{}_{\eta} \xi\dagg\xi 
  = c \log (1+|\ph|^2) \;.
\eeq 

\medskip
2) $G_{N,M} = SU(N)/[SU(N-M) \times SU(M) \times U(1)]$.\\
This is called the (complex) Grassmann manifold. 
We take the $Y$-charge as $Y \equiv -Q_1$ with 
$Q_1 = {\rm diag.}\,(\underbrace{M,\cdots, M}_{N-M}, 
\underbrace{M-N,\cdots,M-N}_M) \in {\cal H}$.
Then complex unbroken and broken generators are given by
\beq
\hat{\cal H} &=& 
  \left(\begin{array}{c|c}
 {\cal SU}(N-M)^{\bf C} & {\bf 0}_{(N-M)\times M} \\ \hline
 {\bf B}_{M\times (N-M)}& {\cal SU}(M)^{\bf C}              
 \end{array}\right)  
 \oplus \cu(1)\bc \,, \non
{\cal G}^{\bf C} - \hat{\cal H} &=& 
 \left(\begin{array}{c|c}
 {\bf 0}_{N-M}           & {\bf P}_{(N-M)\times M} \\ \hline
 {\bf 0}_{M\times (N-M)} & {\bf 0}_M 
 \end{array}\right) \,,
\eeq
respectively, with $\cu(1)\bc$ generated by $Q_1$. 
The complex coset representative is obtained as
\beq
 \ph \cdot Z = 
 \left(\begin{array}{cc}
              {\bf 0}_{N-M} & \ph \cr
  {\bf 0}_{M\times (N-M)}   & {\bf 0}_M \end{array} \right) \;,\hs{5} 
 \xi = \left(\begin{array}{cc}
  {\bf 1}_{N-M} & \ph \cr
  {\bf 0}_{M\times (N-M)}  & {\bf 1}_M \end{array} \right) 
      \label{GNM-coset}
\eeq
with $\ph$ an $N-M$ by $M$ matrix.
The projection operator is $\eta = {\rm diag.}\,
(\underbrace{0,\cdots,0}_{N-M},\underbrace{1,\cdots,1}_M)$ 
and the K\"ahler potential is 
\beq
 K = c \log \det{}_{\eta} \xi\dagg\xi 
  = c \log \det{}_{M\times M} ({\bf 1}_M + \ph\dagg\ph) \;.
\eeq

\medskip
3) $SU(N)/[SU(N-2)\times U(1)^2]$.\\
This coset space allows two inequivalent 
complex structures~\cite{BN,BE}. 
The $U(1)$ generators in $H$ are 
\beq
 Q_1 \equiv {\rm diag.} 
            (\underbrace{1,\cdots,1}_{N-2},-N+2 , 0)\;, \hs{5}
 Q_2 \equiv {\rm diag.} 
            (\underbrace{1,\cdots,1}_{N-2}, 0, -N+2) \;.
 \label{twoU(1)s}
\eeq
Two complex structures 
$\Omega_{\rm I}$ and $\Omega_{\rm II}$ denoted in \cite{BN} 
are represented 
for instance by $Y_{\rm I} \equiv - Q_2$ 
and $Y_{\rm II} \equiv Q_1 -Q_2$, respectively. 
According to these $Y$-charges, 
complex unbroken and broken generators are given by
\beq
\hat {\cal H}_{\rm I} 
 = 
\left( \begin{array}{c|c|c}
                          & 0       & 0 \cr
    \csu (N-2)\bc         & \vdots  & \vdots  \cr 
                          & 0       & 0 \cr \hline
 {\rm B} \;\;\; \cdots \;\;\; {\rm B} & 0     & 0 \cr \hline
 {\rm B} \;\;\; \cdots \;\;\; {\rm B} &{\rm B}& 0 
 \end{array}\right) \oplus 2 \cu (1)\bc 
 ,\hs{2}
({\cal G}^{\bf C} -\hat {\cal H})_{\rm I} 
 = 
\left( \begin{array}{ccc|c|c}
        &        &        & {\rm P} & {\rm P} \cr
  &{\bf 0}_{N-2} &        & \vdots  & \vdots  \cr 
        &        &        & {\rm P} & {\rm P} \cr \hline
      0 & \cdots &  0     & 0       & {\rm P} \cr \hline
      0 & \cdots &  0     & 0       & 0 
 \end{array}\right) 
 \hs{2} \nonumber
\eeq
for $\Omega_{\rm I}$ and
\beq 
\hat {\cal H}_{\rm II} 
 = 
\left( \begin{array}{c|c|c}
                                & {\rm B} & 0 \cr
    \csu (N-2)\bc               & \vdots  & \vdots  \cr 
                                & {\rm B} & 0 \cr \hline
       0 \;\;\; \cdots \;\;\; 0 & 0       & 0 \cr \hline
 {\rm B} \;\;\; \cdots \;\;\; {\rm B} &{\rm B}& 0 
 \end{array}\right) \oplus 2 \cu (1)\bc 
 ,\hs{2}
({\cal G}^{\bf C} -\hat {\cal H})_{\rm II} =
\left( \begin{array}{ccc|c|c}
        &        &        & 0       & {\rm P} \cr
  &{\bf 0}_{N-2} &        & \vdots  & \vdots  \cr 
        &        &        & 0       & {\rm P} \cr \hline
 {\rm P}& \cdots &{\rm P} & 0       & {\rm P} \cr \hline
      0 & \cdots &  0     & 0       & 0 
 \end{array}\right) \hs{3}\nonumber  
\eeq
for $\Omega_{\rm II}$. 
For both cases two $U(1)$ generators are 
given by Eq.~(\ref{twoU(1)s}).
The coset representative for $\Omega_{\rm I}$ is calculated as
\beq
 (\ph \cdot Z )_{\rm I}
= \left( \begin{array}{c|c|c}
   {\bf 0}_{N-2}  & \chi & \phi  \cr \hline
         {\bf 0}  & 0    & \kappa \cr \hline
         {\bf 0}  & 0    & 0 
 \end{array}\right)   ,\hs{5}
\xi_{\rm I} =  \left( \begin{array}{c|c|c}
   {\bf 1}_{N-2}  & \chi & \phi + \1{2} \kappa \chi  \cr \hline
         {\bf 0}  & 1    & \kappa \cr \hline
         {\bf 0}  & 0    & 1 
 \end{array}\right) ,   
\eeq
with $\chi$ and $\phi$ belonging to ${\bf N-2}$ of $SU(N-2)$. 
The coset representative for $\Omega_{\rm II}$ is
\beq
 (\ph \cdot Z)_{\rm II}
 = \left( \begin{array}{c|c|c}
 {\bf 0}_{N-2} & {\bf 0} & \phi   \cr \hline
        \chi^T & 0       & \kappa \cr \hline 
      {\bf 0}  & 0       & 0    
 \end{array}\right) , \hs{5}
\xi_{\rm II} 
 = \left( \begin{array}{c|c|c}
 {\bf 1}_{N-2} & {\bf 0} & \phi   \cr \hline
        \chi^T & 1       & \kappa + \1{2} \chi \cdot \phi \cr \hline
      {\bf 0}  & 0       & 1    
 \end{array}\right) \, .
\eeq
with $\phi$ ($\chi$) belonging to ${\bf N-2}$ ($\overline{{\bf N-2}}$).
The projection operators are given by
\beq
 && (\eta_1)_{\rm I} = \diag (\underbrace{0,\cdots,0}_{N-2},0,1) \;,\hs{5} 
 (\eta_2)_{\rm I} = \diag (\underbrace{0,\cdots,0}_{N-2},1,1) \, ,\\
 && (\eta_1)_{\rm II} = \diag (\underbrace{0,\cdots,0}_{N-2},0,1) \;,\hs{5} 
 (\eta_2)_{\rm II} = \diag (\underbrace{1,\cdots,1}_{N-2},0,1) \,,
\eeq
and the K\"ahler potentials can be calculated as~\cite{BE}
\beq
&& K_{\rm I} =
   c_1 \log \left( 1 + |\kappa|^2 
                 + \left|\phi + \1{2} \kappa \chi\right|^2\right) \non
&& \hs{10}
 + c_2 \log \left( 1 + |\chi|^2 
          + \left| \phi - \1{2} \kappa \chi\right|^2
          + |\chi|^2|\phi|^2 - |\chi\dagg\phi|^2  \right) \,,\\
&& K_{\rm II} 
= c_1 \log \left( 1 + |\phi|^2 + 
  \left|\kappa + \1{2} {\chi \cdot \phi} \right|^2 \right) 
 + c_2 \log \left( 1 + |\chi|^2 + 
  \left|\kappa - \1{2} {\chi \cdot \phi} \right|^2 \right) , \;\;\;\;
 \nonumber \label{omegaII} 
\eeq
for $\Omega_{\rm I}$ and $\Omega_{\rm II}$, respectively.
The one formulated by the auxiliary field method is the second one.

\medskip
4) $SU(N)/[SU(N-M-L) \times SU(M) \times SU(L) \times U(1)^2]$\\
Two $U(1)$ generators in ${\cal H}$ are given by~\footnote{
We have chosen these charges in a different way from 
those in Refs.~\cite{Ku,HKN3}.
}
\beq
 && Q_1 \equiv \diag (\underbrace{L,\cdots,L}_{N-M-L},
  \underbrace{-N+M+L,\cdots, -N+M+L}_L,
  \underbrace{0,\cdots,0}_M) \,,\non
 && Q_2 \equiv \diag (\underbrace{M,\cdots,M}_{N-M-L},
  \underbrace{0,\cdots,0}_L,
  \underbrace{-N+M+L,\cdots, -N+M+L}_M) \;.
\eeq
As the same with the last example, 
the $Y$-charge  
can be taken for instance as $Y_{\rm I} \equiv - Q_2$ 
or $Y_{\rm II} \equiv M Q_1 - L Q_2$ for 
the complex structure $\Omega_{\rm I}$ or $\Omega_{\rm II}$, 
respectively.
Complex unbroken and broken generators are given by
\beq
&&
 \hat {\cal H}_{\rm I}  
= \left( \begin{array}{c|c|c}
 \csu(N-M-L)\bc &{\bf 0}_{(N-M-L)\times L}&{\bf 0}_{(N-M-L)\times M} 
                                                    \cr \hline
 {\bf B}_{L\times(N-M-L)}&\csu(L)\bc    &{\bf 0}_{L\times M} \cr \hline
 {\bf B}_{M\times(N-M-L)}&{\bf B}_{M\times L}& \csu(M)\bc
 \end{array}\right)  
  \oplus 2 \cu(1)\bc \;, \hs{5}  \non
&& ({\cal G}^{\bf C} - \hat {\cal H})_{\rm I} 
 = \left( \begin{array}{c|c|c}
  {\bf 0}_{N-M-L} & {\bf P}_{(N-M-L)\times L} 
                  & {\bf P}_{(N-M-L)\times M} \cr \hline
 {\bf 0}_{L\times (N-M-L)} & {\bf 0}_{L} & {\bf P}_{L\times M} \cr \hline
 {\bf 0}_{M\times (N-M-L)} & {\bf 0}_{M\times L} & {\bf 0}_{M} 
 \end{array}\right)   \;, 
\eeq
for $\Omega_{\rm I}$ and
\beq
&& 
 \hat {\cal H}_{\rm II} 
= \left( \begin{array}{c|c|c}
 \csu(N-M-L)\bc &{\bf B}_{(N-M-L)\times L} & {\bf 0}_{(N-M-L)\times M} 
                                                   \cr \hline
 {\bf 0}_{L\times (N-M-L)} & \csu(L)\bc & {\bf 0}_{L\times M} \cr \hline
 {\bf B}_{M\times(N-M-L)}&{\bf B}_{M\times L}    & \csu(M)\bc
 \end{array}\right)  
  \oplus 2 \cu(1)\bc  \;, \hs{5} \non
&&
({\cal G}^{\bf C} -\hat {\cal H})_{\rm II} 
 = \left( \begin{array}{c|c|c}
 {\bf 0}_{N-M-L} &{\bf 0}_{(N-M-L)\times L} & {\bf P}_{(N-M-L)\times M} 
                                                    \cr \hline
 {\bf P}_{L\times (N-M-L)}&{\bf 0}_L& {\bf P}_{L\times M} \cr \hline
 {\bf 0}_{M\times (N-M-L)}&{\bf 0}_{M\times L}&{\bf 0}_M 
 \end{array}\right)  
\eeq
for $\Omega_{\rm II}$.
The complex coset representatives can be calculated as
\beq
 \hs{-20}
&& (\ph \cdot Z)_{\rm I}
 =  \left(
  \begin{array}{ccc}
   {\bf 0}_{N-M-L} & \chi      & \ph \\
           {\bf 0} & {\bf 0}_L & \kappa \\
           {\bf 0} & {\bf 0}   & {\bf 0}_M
  \end{array} \right) , \ \
\xi_{\rm I} =  \left(
\begin{array}{ccc}
 {\bf 1}_{{N-M-L}} & \chi      & \ph + \1{2} \chi\kappa \\
           {\bf 0} & {\bf 1}_L & \kappa \\
           {\bf 0} & {\bf 0}   & {\bf 1}_M
\end{array} \right) , \\
&& (\ph \cdot Z)_{\rm II}
 =  \left(
 \begin{array}{ccc}
  {\bf 0}_{N-M-L} & {\bf 0}   & \ph \\
           \chi^T & {\bf 0}_L & \kappa \\
          {\bf 0} & {\bf 0}   & {\bf 0}_M
\end{array} \right),  \ \
\xi_{\rm II} =  \left(
\begin{array}{ccc}
{\bf 1}_{{N-M-L}} & {\bf 0}   & \ph \\
           \chi^T & {\bf 1}_L & \kappa + \1{2} \chi^T\ph \\
          {\bf 0} & {\bf 0}   & {\bf 1}_M
\end{array} \right)  \label{xi-II} \hs{10}
\eeq
for $\Omega_{\rm I}$ and $\Omega_{\rm II}$, respectively.
The projection operators are given by
\beq
 && (\eta_1)_{\rm I} 
  = \diag (\underbrace{0,\cdots,0}_{N-M-L},
           \underbrace{0,\cdots,0}_{L},
           \underbrace{1,\cdots,1}_{M}) \;,\non 
 &&(\eta_2)_{\rm I} 
  = \diag (\underbrace{0,\cdots,0}_{N-M-L},
           \underbrace{1,\cdots,1}_{L},,
           \underbrace{1,\cdots,1}_{M}) \, 
\eeq
for $\Omega_{\rm I}$ and 
\beq
 && (\eta_1)_{\rm II} 
  = \diag (\underbrace{0,\cdots,0}_{N-M-L},
           \underbrace{0,\cdots,0}_{L},,
           \underbrace{1,\cdots,1}_{M}) \;,\non
 && (\eta_2)_{\rm II} 
  = \diag (\underbrace{1,\cdots,1}_{N-M-L},
           \underbrace{0,\cdots,0}_{L},,
           \underbrace{1,\cdots,1}_{M}) \,
\eeq
for $\Omega_{\rm II}$.
Using these projection operators, 
the K\"ahler potentials are calculated as
\beq
 &&
 K_{\rm I} 
 = c_1 \log \det \left[ {\bf 1}_M + \kappa\dagg\kappa 
        + \left( \ph\dagg + \1{2} \kappa\dagg\chi\dagg \right) 
          \left( \ph + \1{2} \chi \kappa \right) \right] \non 
 && \hs{2}
 + c_2 \log \det
  \left(
  \begin{array}{cc}
    {\bf 1}_{L} + \chi\dagg\chi 
      & \kappa + \chi\dagg(\ph + \1{2}\chi\kappa)\cr
    \kappa\dagg + (\ph\dagg + \1{2}\kappa\dagg\chi\dagg)\chi
      & {\bf 1}_M + \kappa\dagg\kappa
          + \left( \ph\dagg + \1{2}\kappa\dagg\chi\dagg \right)
            \left( \ph + \1{2}\chi\kappa \right) 
  \end{array} 
  \right),  \non
 &&
 K_{\rm II}
 = c_1 \log \det \left[ {\bf 1}_M + \ph\dagg\ph 
          + \left( \kappa + \1{2} \ph\dagg \chi^* \right)
            \left( \kappa + \1{2} \chi^T \ph \right) \right] \\ 
 && \hs{2}
 + c_2 \log \det
  \left(
  \begin{array}{cc}
    {\bf 1}_{N-M-L} + \chi^*\chi^T 
      & \ph + \chi^*(\kappa + \1{2}\chi^T\ph)\cr
    \ph\dagg + (\kappa\dagg + \1{2}\ph\dagg\chi^*)\chi^T
      & {\bf 1}_M + \ph\dagg\ph
          + \left( \kappa\dagg + \1{2} \ph\dagg\chi^* \right)
            \left( \kappa + \1{2} \chi^T\ph \right) 
  \end{array} 
  \right),  \nonumber
\eeq
for $\Omega_{\rm I}$ and $\Omega_{\rm II}$, respectively.

The second one should coincide with 
the K\"ahler potential (\ref{result2}) up to 
a holomorphic coordinate transformation and 
a K\"ahler transformation, but we are unable to 
show their equivalence.


\section{Some geometric structures}
In this appendix, 
we discuss more geometric structures of the manifolds 
presented in this paper; 
their bundle structures and 
their relation with hyper-K\"ahler manifolds 
and the Calabi-Yau manifolds of 
cohomogeneity one~\cite{HKN1}--\cite{Ni2}.

First we discuss the bundle structures 
using gauging/ungauging technique~\cite{HKN2}.
If we gauge a part of isometry $I \subset G$ 
on $\hat {\cal M}$ introducing vector superfields $V$ 
and then integrate $V$ out, 
we obtain a K\"ahler quotient manifold 
${\cal M} = \hat {\cal M}/I\bc$.
On the other hand, if we ``ungauge'' $I$ by freezing $V$ 
in a K\"ahler quotient formulation of ${\cal M}$,  
we obtain $\hat {\cal M}$.
In the case of $I = U(1)$ [$I=U(M)$], 
$\hat {\cal M}$ can be regarded as 
a complex line (${\bf C}^M$-) bundle over ${\cal M}$.
By applying this technique to 
Hermitian symmetric spaces (HSS) formulated 
as gauge theories~\cite{HN1}, 
canonical complex line bundles over HSS are 
constructed in \cite{HKN2}.

We now discuss relations between manifolds in this paper
with other manifolds.
First let us consider $SU(N)/[SU(N-2)\times U(1)^2]$.
If we put $V \equiv  V_1 =  - V_2$ 
(or freezing $V_1 + V_2$) 
in the Lagrangian (\ref{Lag1}) 
we get 
\beq
 {\cal L} = \int d^4 \theta (e^{V} \phi_1{}\dagg\phi_1 
 + e^{- V}\phi_2{}\dagg\phi_2 - c V )
 + \left(\int d^2\theta \lam \phi_1 \cdot \phi_2 + {\rm c.c.}\right)\;,
 \label{HKquotient}
\eeq
with $c \equiv c_2 -c_1$. 
SUSY is enhanced to $D=2$, ${\cal N}=4$ ($D=4$, ${\cal N}=2$) SUSY.
For this SUSY, target space must be hyper-K\"ahler (HK)~\cite{AF}.
Actually, Eq.~(\ref{HKquotient})  
is the HK quotient construction~\cite{LR,HKLR} for 
the HK Calabi metric~\cite{Ca} 
on the cotangent bundle over ${\bf C}P^{N-1}$, 
$T^* {\bf C}P^{N-1}$~\cite{HKQ-CPN,ANNS}. 
Therefore $SU(N)/[SU(N-2)\times U(1)^2]$ is a $U(1)$ K\"ahler 
quotient of $T^* {\bf C}P^{N-1}$, and 
the latter is 
a complex line bundle over the former.
$T^* {\bf C}P^{N-1}$ is the only one HK manifold 
of cohomogeneity one~\cite{DS}.
In our method, 
the cohomogeneity of $T^* {\bf C}P^{N-1}$ is 
easily found to be one, 
because we construct it from a compact manifold 
by freezing one gauge degree of freedom.

Next let us consider non-Abelian case of $M=L$: 
$SU(N)/[SU(N-2M) \times SU(M)^2\times U(1)^2]$.
If we put $V \equiv V_1 = -V_2^T$ 
in the Lagrangian (\ref{Lag2}) 
with $M=L$, 
we get 
\beq
 {\cal L} = \int d^4\theta \left[ \tr(\Phi\dagg\Phi e^{V})
 + \tr(\Psi^T \Psi^* e^{- V}) - c\, \tr V \right] 
 + \left[ \int d^2\theta \, \tr (\lam \Psi^T \Phi) + {\rm c.c.} \right] , 
 \label{T*GNM}
\eeq
with $c \equiv c_2 -c_1$. 
As the same as the above discussions, 
this is the HK quotient construction for 
the Lindstr\"om-Ro\v{c}ek metric on $T^* G_{N,M}$~\cite{LR,ANS}.
(Putting $\Psi =0$ in the Lagrangian (\ref{T*GNM}) 
we obtain the Lagrangian (\ref{GNM}) for $G_{N,M}$, 
and therefore we find this bundle structure.)
Thus $SU(N)/[SU(N-2M)\times U(M)^2]$ is 
a $U(M)$ K\"ahler quotient of $T^* G_{N,M}$, and 
the latter is a ${\bf C}^M$-bundle over the former.

\medskip
Before closing this appendix, we discuss 
possibility to construct a new Calabi-Yau (CY) metric 
of cohomogeneity one~\cite{HKN1}--\cite{Ni2}. 
If we freeze out $V_1 + V_2$ in $SU(N)/[SU(N-2)\times U(1)^2]$ 
starting from the most general Lagrangian, 
we obtain
\beq
 {\cal L} = \int d^4 \theta 
  [e^{V} \phi_1{}\dagg\phi_1 + f( e^{- V}\phi_2{}\dagg\phi_2) - c V ]
 + \left(\int d^2\theta \lam \phi_1 \cdot \phi_2 + {\rm c.c.}\right)\;, 
\eeq
with $f$ an arbitrary function.\footnote{
We can take this K\"ahler potential as the most general one 
instead of 
$f(e^{V} \phi_1{}\dagg\phi_1, e^{- V}\phi_2{}\dagg\phi_2)$ 
because it can be shown that one variable can 
be linearized when $V$ is integrated~\cite{HN2}.
}
This is a deformation of HK metric (\ref{HKquotient}) 
preserving only the K\"ahler structure.
If we freeze out $a V_1 + b V_2$ (with $a,b \in {\bf R}$) 
in $SU(N)/[SU(N-2)\times U(1)^2]$, we get 
\beq
 {\cal L} = \int d^4 \theta 
 [e^{V} \phi_1{}\dagg\phi_1 
 + f (e^{q V}\phi_2{}\dagg\phi_2) - c V ]
 + \left(\int d^2\theta \lam \phi_1 \cdot \phi_2 + {\rm c.c.}\right)\;, 
\eeq
with $q \equiv - a/b$ being 
the relative charge of the remained $U(1)$ gauge 
for $\phi_1$ and $\phi_2$, 
and $c \equiv c_1 + q c_2$.
This is the Lagrangian suggested in \cite{Ni2} 
[(A.2) in Appendix]
as a generalization of the construction of 
a CY metric using matter coupling in the ${\bf C}P^N$ model.
Integrating $V$ and $\lam$ we obtain 
\beq
 {\cal L} = \int d^4 \theta 
 \left[ \log \phi_1{}\dagg\phi_1 
 + h ( (\phi_1{}\dagg\phi_1)^{-q} \phi_2{}\dagg\phi_2 ) \right]
\eeq
with the constraint $\phi_1 \cdot \phi_2 = 0$ and 
some function $h$ related with $f$.
In the case of $q=0$, 
the CY metric was obtained in \cite{Ni2} 
solving the Ricci-flat condition for $h$.
The case of $q=-1$ corresponds to the HK Calabi metric.
If we solve it for general $q$, 
we will be able to obtain the most general CY metric on 
the line bundle over $SU(N)/[SU(N-2)\times U(1)^2]$, if it exists,  
which is cohomogeneity one and can be locally written as
\beq 
 {\bf R} \times SU(N)/[SU(N-2)\times U(1)]\,,
\eeq
where freedom to embed $U(1) \subset H$ 
into $SU(N)$ corresponds to $q$.
The holonomy structure for 
the case of $N=3$, ${\bf R} \times SU(3)/U(1)$, 
was discussed in detail in \cite{KY} 
including $Spin(7)$ holonomy.

\end{appendix}



\end{document}